\documentclass[aps,prl,twocolumn,amsmath,amssymb,superscriptaddress,longbibliography]{revtex4-1}
\usepackage[english]{babel}

\usepackage[utf8]{inputenc}
\usepackage[T1]{fontenc}

\usepackage{bbm}
\usepackage{amsmath}
\usepackage{verbatim}
\usepackage{graphicx}
\usepackage{times}
\usepackage{amssymb}
\usepackage{epsfig}
\usepackage{graphicx}
\usepackage{bm}
\usepackage{txfonts}
\usepackage{dsfont}
\usepackage{color}

\usepackage{subfigure}

\newcommand{\ket}[1]{|#1\rangle}
\newcommand{\bra}[1]{\langle#1|}

\begin{document}

\selectlanguage{english}

\title{Sub-Poissonian Phonon Lasing in Three-Mode Optomechanics}

\author{Niels L\"orch}
\affiliation{Institut f\"{u}r Gravitationsphysik, Leibniz Universit\"{a}t Hannover and \\ Max-Planck-Institut f\"{u}r Gravitationsphysik (Albert-Einstein-Institut), Callinstra\ss{}e 38, 30167 Hannover, Germany}
\affiliation{Institut f\"ur Theoretische Physik, Leibniz Universit\"at Hannover, Appelstra\ss{}e 2, 30167 Hannover, Germany}

\author{Klemens~Hammerer}
\affiliation{Institut f\"{u}r Gravitationsphysik, Leibniz Universit\"{a}t Hannover and \\ Max-Planck-Institut f\"{u}r Gravitationsphysik (Albert-Einstein-Institut), Callinstra\ss{}e 38, 30167 Hannover, Germany}
\affiliation{Institut f\"ur Theoretische Physik, Leibniz Universit\"at Hannover, Appelstra\ss{}e 2, 30167 Hannover, Germany}

\date{\today}

\begin{abstract}
We propose to use the resonant enhancement of the parametric instability in an optomechanical system of two optical modes coupled to a mechanical oscillator to prepare mechanical states with sub-Poissonian phonon statistics. Strong single photon coupling is not required. The requirements regarding sideband resolution, circulating cavity power and environmental temperature are in reach with state of the art parameters of optomechanical crystals. Phonon antibunching can be verfied in a Hanburry-Brown-Twiss measurement on the output field of the optomechanical cavity.
\end{abstract}


\maketitle

\paragraph{Introduction}

Optomechanical experiments, where light resonators are coupled to mechanical oscillators
\cite{Aspelmeyer2014,Aspelmeyer2014a}, are achieving increasingly good control of macroscopic objects on the quantum level: Milestones such as cooling the motion of these oscillators to their quantum ground state \cite{Teufel2011,Chan2011a}, coherent transfer of information between light and mirror \cite{Verhagen2012, Palomaki2013}, observation of radiation pressure
shot noise on the oscillator \cite{Murch2008a,Purdy2013a}, as well as entanglement
between the light field and the mirrors \cite{Palomaki2013a} have been achieved in recent years.

The phonon analogue of a laser, which is realized using the optical cavity as the gain medium to excite coherent oscillations of the mechanical oscillator has been demonstrated in \cite{Carmon2005,Tomes2009,Grudinin2009,Grudinin2010,Bahl2011,Anetsberger2011,Zaitsev2011,Suchoi,Cohen2014}, and  its phonon statistics has been mapped out via a Hanburry-Brown-Twiss measurement on the sideband-photons emitted from the optomechanical cavity \cite{Cohen2014}.  Theoretical work suggests that it is possible to prepare a state with quantum signatures in the phonon statistics such as phonon antibunching and even negative Wigner density \cite{Rodrigues2010,Armour2012a,Qian2012,Nation2013,Lorch2014,Nation2015}. However, the requirements on system parameters to see phonon antibunching scale unfavorably,
so that sub-Poissonian phonon statistics has eluded experimental observation.

In this article we propose to make use of the enhanced optomechanical nonlinearity \cite{Safavi-Naeini2011,Ludwig2012,Xu2014} of a setup with two optical modes
to overcome this difficulty and prepare phonon laser states featuring antibunching in steady state with state of the art optomechanical crystals. The enhanced nonlinearity has been discussed in the context of detectors for phonons or photons \cite{Ludwig2012}, quantum memory \cite{Komar2013}, and to improve \cite{Xu2014} the parameters of mechanically induced photon antibunching \cite{Rabl2011,Nunnenkamp2011}. In the context of the phonon laser transition the enhanced optomechanical instability with two optical modes has been anticipated as a possible complication for gravitational wave detectors \cite{Braginsky2001}, and has been studied experimentally  \cite{Tomes2009,Grudinin2009,Grudinin2010,Bahl2011,Anetsberger2011,Chen2014} and theoretically \cite{Wu2013,Danilishin2014,Ju2014,Wang2014a} in the classical regime.
Here we show for the first time that one can detect quantum signatures in the phonon lasing of such a three-mode system. In particular phonon antibunching and, with more demanding system requirements, negative mechanical Wigner density can be prepared in steady state.

\paragraph{Terminology for phonon statistics}

Denoting the phonon number $\hat n=c^\dagger c$, its statistics is characterized by the Fano factor $F=\langle \Delta \hat n^2 \rangle/\langle  \hat n \rangle$, and the second order coherence function $g^{(2)}(t)$ at time $t=0$
\begin{align}
g^{(2)}(0)=\frac{\langle c^\dagger c^\dagger cc\rangle}{\langle \hat n \rangle^2}=1+ (F-1)/\langle \hat n \rangle, \label{g2relF}
\end{align}
which gives information on the temporal correlations of the phonons. ( $g^{(2)}(0)>1$ and $g^{(2)}(0)<1$ corresponding to bunching and anti-bunching respectively \cite{Kimble1977}.)
The Fano factor $F$ can be inferred from $g^{(2)}(0)$ through \eqref{g2relF}, and $F$ smaller/greater than 1 indicates sub/super-Poissonian statistics.
In \cite{Cohen2014} $g^{(2)}(0) \approx 1$ was achieved, verifying the coherent nature of the mechanical oscillations in their setup. For comparison, the Poissonian statistics of a (classical) coherent state imply $F=1$ and $g^{(2)}(0)=1$, while a thermal state would have $g^{(2)}(0)=2$.

\begin{figure}[t]
  \includegraphics[width=\columnwidth]{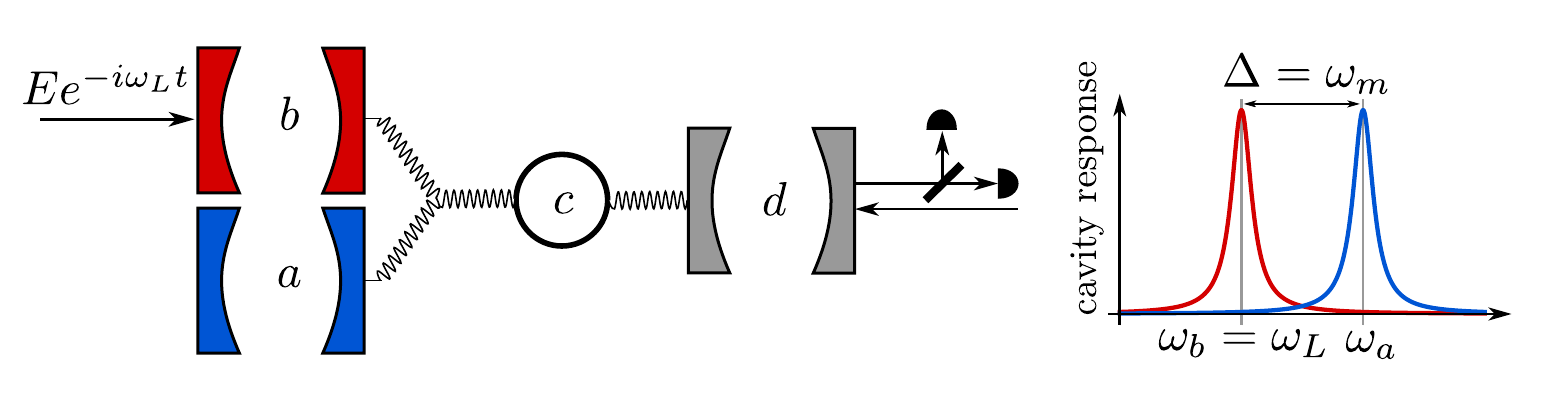}
	\caption{left: Two optical modes $a$ and $b$ are coupled to a mechanical mode $c$. The $b$-mode is resonantly driven by a laser of strength $E$ and frequency $\omega_L=\omega_b$. The $a$-mode is detuned from $b$ by $\Delta=\omega_m$, the mechanical resonance frequency, as depicted in the plot on the right. The nonlinear interaction of the three modes $a$, $b$, and $c$ gives rise to optomechanical limit cycles with strongly sub-Poissonian phonon number statistics. A third optical mode $d$ can be used to reduce the effective temperature of the mechanical oscillator's bath and to read out the phonon statistics in a Hanburry-Brown-Twiss measurement. }
	\label{SETUP}
\end{figure}

\paragraph{Description of the system}

We study the optomechanical setup depicted in Fig.~ \ref{SETUP}. Two optical modes $a$ and $b$ couple to a mechanical mode $c$ via the three-mode interaction Hamiltonian
$
V=g_0 (a b^\dagger + a^\dagger b) (c+ c^\dagger),
$where $g_0$ is the single photon optomechanical coupling strength and $a,b,c$ are the lowering operators of the different modes.  Such an interaction has been implemented in Refs. \cite{Tomes2009,Grudinin2009,Grudinin2010,Bahl2011,Anetsberger2011,Chen2014}. The optical mode $b$ is resonantly driven with a laser of power $P$ , which we parametrize with
 $E=\sqrt{\kappa P/\hbar\omega_b}$ ($\kappa$ is the cavity line width, and $\omega_b$ the resonance frequency of mode $b$). The other optical mode $a$ is detuned with respect to cavity mode $b$ and the driving laser by $\Delta$, and the mechanical frequency is $\omega_m$, so that the Hamiltonian in a rotating frame for both cavities with frequency $\omega_b$ is $H=H_0+V+iE( b^\dagger-b)$ with $H_0=\omega_m c^\dagger c-\Delta a^\dagger a $.

Depending on the sign of the laser detuning, the laser either cools the mechanical mode ($\Delta<0$)
, or gives rise to self-induced mechanical oscillations ($\Delta>0$). In the latter regime the intrinsic nonlinearity of the three-mode optomechanical interaction $V$ stabilizes the mechanical oscillation at a finite amplitude \cite{Chen2014}. We choose a detuning $\Delta=\omega_m$ between the two cavities which corresponds to a resonant excitation of optomechanical limit cycles. In an interaction picture with respect to $H_0$ the Hamiltonian is
\begin{align}
H_I&=iE( b^\dagger-b)+g_0 (a b^\dagger c + a^\dagger b c^\dagger). \label{totH}
\end{align}
We neglected here fast oscillating terms $e^{2 i \omega_m t}g_0 a b^\dagger c^\dagger  +h.c.$, assuming a cavity decay rate of $\kappa \ll \omega_m$ for both cavities (the corrections are of order $\kappa^2/\omega_m^2$, i.e. negligible for typical optomechanical crystals.).  In the framework of Langevin equations the system dynamics is then described by
\begin{align}
&\dot a=-i g_0 b c^\dagger -\tfrac \kappa 2 a +\sqrt \kappa a_{in}  \label{Langevin1},  \\
&\dot b=-i g_0 a c  -\tfrac \kappa 2 b +E +\sqrt \kappa b_{in}  \label{Langevin2},  \\
& \dot c = -i g_0 a^\dagger b -\tfrac \gamma 2 c +\sqrt \gamma c_{in}, \label{Langevin}
\end{align}
where $\langle a_{in}(t) a_{in}^\dagger(t') \rangle=\langle b_{in}(t) b_{in}^\dagger(t') \rangle= \delta(t-t')$ and $\langle c_{in}(t) c_{in}^\dagger(t') \rangle= (1+\bar n) \delta(t-t')$ are the two-time correlation functions of the Langevin noise forces. We assumed energy decay of the mechanical oscillator at rate $\gamma$, due to coupling to a thermal thermal bath with mean occupation $\bar n$.   We adopt the convention from the review \cite{Aspelmeyer2014} that $\kappa$ and $\gamma$ are energy decay rates. Correspondingly, amplitudes decay at $\kappa/2$ and $\gamma/2$.

\begin{comment}

\begin{figure}[t]
  \includegraphics[width=0.5\textwidth]{nonlinDamping.pdf}
	\caption{b) Optically mediated (anti)damping $\gamma_{\mathrm{opt}}$ (bold line) as a function of mechanical amplitude $\zeta$ according to equation \eqref{gamma}. The steady state $\zeta_0$ is reached where $\gamma_{\mathrm{opt}}(\zeta_0)=-\gamma$.  As an example with gain $\mathcal R=10$ we draw also $-\gamma$ (dashed line) so that the $\zeta_0$ is at the intersection of both lines. } \label{damping}
\end{figure}

\begin{figure}[t]
  \includegraphics[width=0.25\textwidth]{squeezing2.pdf}
	\caption{schematic drawing of the mechanical oscillator in phase space. The $X$ quadrature relates to the phase of the oscillator and the $Y$ quadrature to its amplitude. } \label{squeezing}
	\label{phasespace}
\end{figure}

\end{comment}

\begin{figure}[t]
  \includegraphics[width=0.45\textwidth]{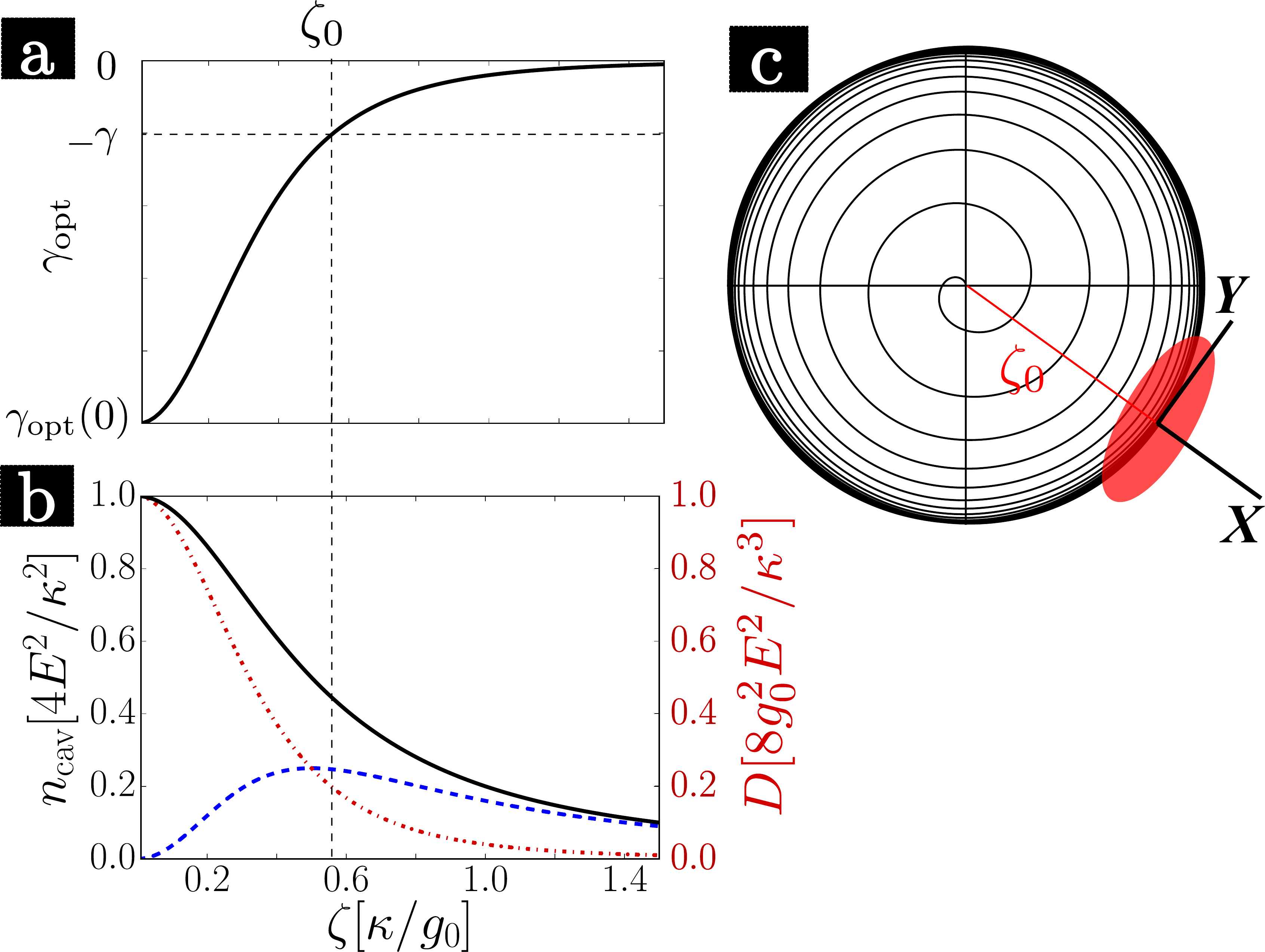}
	\caption{ (a) Optically mediated (anti)damping $\gamma_{\mathrm{opt}}(\zeta)$ (bold line) as a function of mechanical amplitude $\zeta$ according to equation \eqref{gamma}. The steady state $\zeta_0$ is reached when $\gamma_{\mathrm{opt}}(\zeta_0)=-\gamma$ (dashed line). (b) Intra cavity photon number $|\alpha|^2$ in mode $a$ (blue dashed line), $|\beta|^2$ in mode $b$ (red dash-dotted line) and total photon number $n_\mathrm{cav}=|\alpha|^2+|\beta|^2$ (black solid line) is plotted as a function of mechanical amplitude $\zeta$ according to equation \eqref{alphabeta}. The optically induced  diffusion $D_{\mathrm{opt}}=g_0^2 \tfrac \kappa 2 (|\alpha|^2+|\beta|^2)/h_\zeta$ of the mechanical oscillator scales exactly like the red dash-dotted line with a scale as given on the right $y$-axis. (c) Schematic phase space trajectory of the mechanical oscillator approaching the limit cycle attractor with amplitude $\zeta_0$. In the co-rotating frame of the oscillator the $X$ quadrature relates to its amplitude and the
$Y$ quadrature to its phase.} \label{phasespace}

\end{figure}

%%%%%%%%%%%%%%%%%%%%%%%%%%%%%%%%

\paragraph{Calculation of classical amplitudes} We express each of the operators $a,b$, and $c$ as a sum of a classical ($\mathds{C}$-number) component and operators describing fluctuations around it, such that $a=\alpha+\delta a$, $b=\beta+\delta b$ and $c=\zeta+\delta c$. Inserting this into the Langevin equations, and considering the $\mathds{C}$-number components only, gives rise to a coupled set of nonlinear equations for the classical cavity amplitudes $\alpha$ and $\beta$, and the (complex) mechanical amplitude $\zeta$. In particular one finds, $\dot{\alpha}=-i g_0 \beta \zeta^* -\tfrac \kappa 2 \alpha$ and $\dot\beta=-i g_0 \alpha \zeta  -\tfrac \kappa 2 \beta +E$.  We assume that the optical amplitudes adiabatically follow the motion of the mechanical oscillator which is equivalent to the conditions $  (\bar n+1) \gamma, g_0 |\alpha|, g_0|\beta| \ll \kappa$. Solving $\dot{\alpha}=\dot\beta=0$ results in the adiabatic solution for the optical amplitudes
\begin{align}
\beta(\zeta,\zeta^*)&=   \frac{E\kappa}{2 {h_\zeta}}, &
\alpha(\zeta,\zeta^*)&=-i  \frac{Eg_0 \zeta^* }{h_\zeta} \label{alphabeta},
\end{align}
where $h_\zeta={g_0^2 |\zeta|^2 + \tfrac14 \kappa^2}$. Inserting these optical amplitudes in the equation of motion for the classical mechanical amplitude results in $\dot \zeta =-\frac 12 (\gamma+\gamma_{\mathrm{opt}}) \zeta$, where the optically mediated (anti)damping is
\begin{align}
 \gamma_{\mathrm{opt}}(\zeta)=-\frac{g_0^2 E^2 \kappa }{h_\zeta^2}, \label{gamma}
\end{align}
cf. Fig.~\ref{phasespace}a.
$\gamma_{\mathrm{opt}}$ is negative for all mechanical amplitudes and its absolute value decreases with increasing amplitude $\zeta$  according to the Lorentzian given by $h_\zeta^2$, approaching 0 for $\zeta \gg \kappa/g_0$.
In agreement with \cite{Chen2014} we define the dimensionless parameter
\begin{align}\label{gainR}
\mathcal R=\frac{|\gamma_{\mathrm{opt}}(0)|}{\gamma}=\frac{16 g_0^2 E^2 }{\kappa^3 \gamma},
 \end{align}
which corresponds to the gain 
of mechanical amplification at zero mechanical amplitude. For $\mathcal R<1$ the total mechanical damping $\gamma+\gamma_{\mathrm{opt}}(0)<0$ is positive for all amplitudes, implying $\zeta=0$ in steady state. Above threshold, $\mathcal R>1$, the steady state ($\dot\zeta=0$) is achieved for a mechanical amplitude $\zeta_0$ such that $\gamma_{\mathrm{opt}}{(\zeta_0)}=\gamma$, cf. Fig.~\ref{phasespace}a. The solution of this nonlinear equation is
\begin{align}
|\zeta_0|^2 =\left(\frac{ \kappa }{2g_0}\right)^2 \left({\sqrt {\mathcal R}-1}\right) \label{delta}.
\end{align}
The solution is unique (up to the oscillator's phase) and fully determined by the gain parameterside $\mathcal R$ and the single-photon strong-coupling parameter $2g_0/\kappa$. It is instructive to contrast this result with the equivalent one for a conventional, two-mode (that is one mechanical and one optical mode) optomechanical system where the mean phonon number of self induced limit cycles scales as the inverse of the much smaller ratio $(g_0/\omega_m)^2$ instead. In view of Eq.~\eqref{g2relF} it is clear that a small oscillation amplitude is advantageous in order to observe strong antibunching
and that the three mode setup improves the signal approximately by a factor of $4(\omega_m/\kappa)^2$. This can be two orders of magnitude for typical system parameters of optomechanical crystals, e.g. $4(\omega_m/\kappa)^2=217$ with $\kappa/2 \pi=500 \mathrm{MHz}$ and $\omega_m/2 \pi=3.68 \mathrm{GHz}$ from \cite{Chan2011a}.

 In the following we will set the arbitrary phase of the limit cycle oscillation to be zero,  $\zeta_0=|\zeta_0|$, without loss of generality. Note also that the cavity amplitudes in Eq.~\eqref{alphabeta} change quite significantly as the mechanical limit cycles develops, cf. Fig.~\ref{phasespace}b, as follows from their enhanced interaction, which detunes the cavity from its input.

\begin{figure}[t]
  \includegraphics[width=0.45\textwidth]{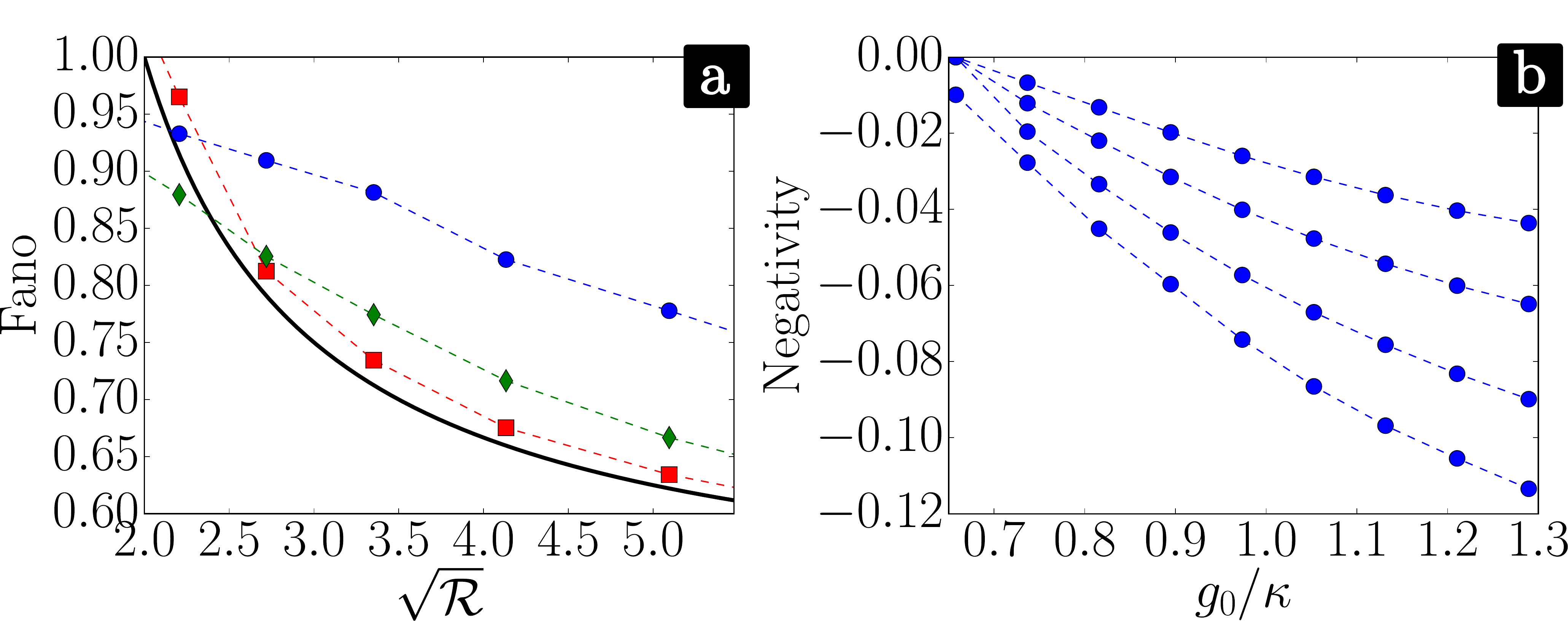}
	\caption{ (a) Comparison of analytical results from Eq. \eqref{Fano} to numerical results for Fano factor $F$ with increasing $g_0/\kappa=0.25,0.5,1.$ (square, diamond, circle). The parameter $g_0 E/\kappa^2=0.04$ is fixed to stay well inside the regime of validity of the adiabatic elimination. In this plot $\bar n=0$ but for finite temperature the agreement of numerics with Eq. \eqref{Fano} is equally good.
(b) Negativity of the Wigner function $W$ of the mechanical oscillator in steady state calculated with QuTiP's steady state solver. We define the negativity as the quotient of the smallest and the largest value of $W$. In this plot the bath occupation is $\bar n=0.25, 0.5, 1, 2$ for the increasing curves. The driving field $E=0.07 \kappa$ is constant in each of these plots, to stay well in the regime of $n_\mathrm{ph} \ll 1$ for numerical simplicity. Each point is optimized over $\mathcal R$ by varying $\gamma$.
	\label{negWig}
}
\end{figure}

\paragraph{Calculation of quantum amplitude noise} The fluctuations $\delta a$, $\delta b$, and $\delta c$ with respect to these classical amplitudes fulfill the linearized Langevin equations
\begin{align}
\delta \dot a&= \left(-\tfrac \kappa 2 \delta a-i g_0 \zeta_0 \delta b\right)  -i g_0 \beta_0 \delta c^\dagger +\sqrt \kappa a_{in},\\
\delta \dot b&= \left(-\tfrac \kappa  2  \delta b-i g_0 \zeta_0 \delta a\right)  -i g_0 \alpha_0 \delta c +\sqrt \kappa b_{in},\\
\delta \dot c&=-\tfrac \gamma 2 \delta c -i g_0(\alpha_0^* \delta b + \beta_0 \delta a^\dagger) +\sqrt \gamma c_{in}, \label{dLangevin}
\end{align}
 where  we consistently dropped all terms of quadratic order in the fluctuations. This approximation is only valid for large enough amplitudes. We also introduce here the shorthand notation $\left(\alpha_0, \beta_0\right) =\left(\alpha(\zeta_0),\beta(\zeta_0)\right)$ for the cavity amplitudes in the developed mechanical limit cycle. The quantum fluctuations of the cavity modes can now be treated in analogy to the classical amplitudes simply by setting $\delta\dot a= \delta\dot b =0$ and solving the resulting algebraic equation.
Inserting the solutions for $\delta a$ and $\delta b$ back into Eq.~ \eqref{dLangevin} gives the dynamics for the mechanical mode $\delta c$. For the canonical mechanical quadratures $X=\left(\delta c+\delta c^\dagger\right)/\sqrt 2$ and $Y=\left(\delta c-\delta c^\dagger\right)/ \sqrt 2i$, cf. Fig.~ \ref{phasespace} c), we get effective Langevin equations
\begin{align}
&\dot X=-\tfrac 12 \Gamma  X +\sqrt D X_N, &\dot Y=\sqrt D Y_N, \label{quadratureEquation}
\end{align}
with damping $\Gamma$, diffusion $D$ and noise forces fulfilling $\langle X_N(t),X_N(t')\rangle=\delta(t-t')$ and $\langle Y_N(t),Y_N(t')\rangle=\delta(t-t')$. Both $\Gamma=\gamma+\Gamma_{\mathrm{opt}}(\zeta)$ and $D=\gamma (\tfrac 1 2+ \bar n)+D_{\mathrm{opt}}(\zeta)$ have an intrinsic mechanical  constant contribution and an optically mediated nonlinear ($\zeta$-dependent) contribution. We find that $D_{\mathrm{opt}}(\zeta)=g_0^2 \tfrac \kappa 2 (|\alpha|^2+|\beta|^2)/h_\zeta$ at the point of the limit cycle is exactly as large as the vacuum contribution of the mechanical bath, i.e. $D_{\mathrm{opt}}(\zeta_0)=\tfrac \gamma 2 $, but $\Gamma_{\mathrm{opt}}(\zeta_0)= \gamma (3-4/\sqrt{\mathcal R})$ can grow up to three times the mechanical damping for large $\mathcal R$.
In total the damping and diffusion depicted in Fig.~ \ref{phasespace} a) are at the limit cycle
\begin{align}
&\Gamma(\zeta_0)= 4 \gamma (1-1/\sqrt{\mathcal R}), &D(\zeta_0) =\gamma (\bar n +1).
\end{align}
  As schematically depicted in Fig.~ \ref{phasespace} c), in our convention the $Y$-quadrature relates to the phase of the mechanical oscillator, which is subjected to undamped diffusion, cf. Eq.~ \eqref{quadratureEquation}. The $X$-quadrature relates to the mechanical amplitude, our focus of interest in this article. In particular for the phonon occupation number $\hat{n}=c^\dagger c$ one finds $\langle \hat n \rangle = \zeta^2 +\mathcal O(\zeta^0)$ and  $\langle \hat n^2 \rangle = \zeta^4+2\zeta^2\langle X^2 \rangle+\mathcal O(\zeta^0) $, such that the Fano factor is $F \simeq 2\langle X^2\rangle $.
 Eq.~ \eqref{quadratureEquation} gives $\langle X^2\rangle= D/\Gamma $ in steady state, i.e. the amplitude variance is determined by the compromise of diffusion and effective damping, yielding for the Fano factor
\begin{align} \label{Fano}
&F 
=\frac{   1}{2}\frac{1+\bar n}{1-1/\sqrt{\mathcal R}}.
\end{align}
This is in excellent agreement with numerical results
shown in Figure \ref{negWig} a)
that were obtained by Monte-Carlo simulation of a master equation equivalent to the exact, nonlinear equations of motion in Eqs.~\eqref{Langevin1} to \eqref{Langevin}. The numerics is further described in the Appendix \cite{Appendix}.

From Eq.~\eqref{Fano} we see that for $\mathcal R \gg 1$ the Fano factor approaches $(1+\bar n)/2$. Therefore, we arrive at the condition $\bar{n}<1$ necessary in order to observe sub-Poissonian phonon statistics. For a cryogenically cooled mechanical oscillator $\bar n=1/(e^{- \hbar \omega_m /k_B T}-1)<1$ can in principle be achieved for a sufficiently high resonance frequency and at low temperature $T$, see \cite{Safavi-Naeini2014,Meenehan2014}. However, in the present case it is possible to take advantage of laser cooling of the mechanical oscillator \cite{WilsonRae2007,Marquardt2007} in order to observe sub-Poissonian statistics.

%\newpage

\begin{figure}[t]
  \includegraphics[width=0.45\textwidth]{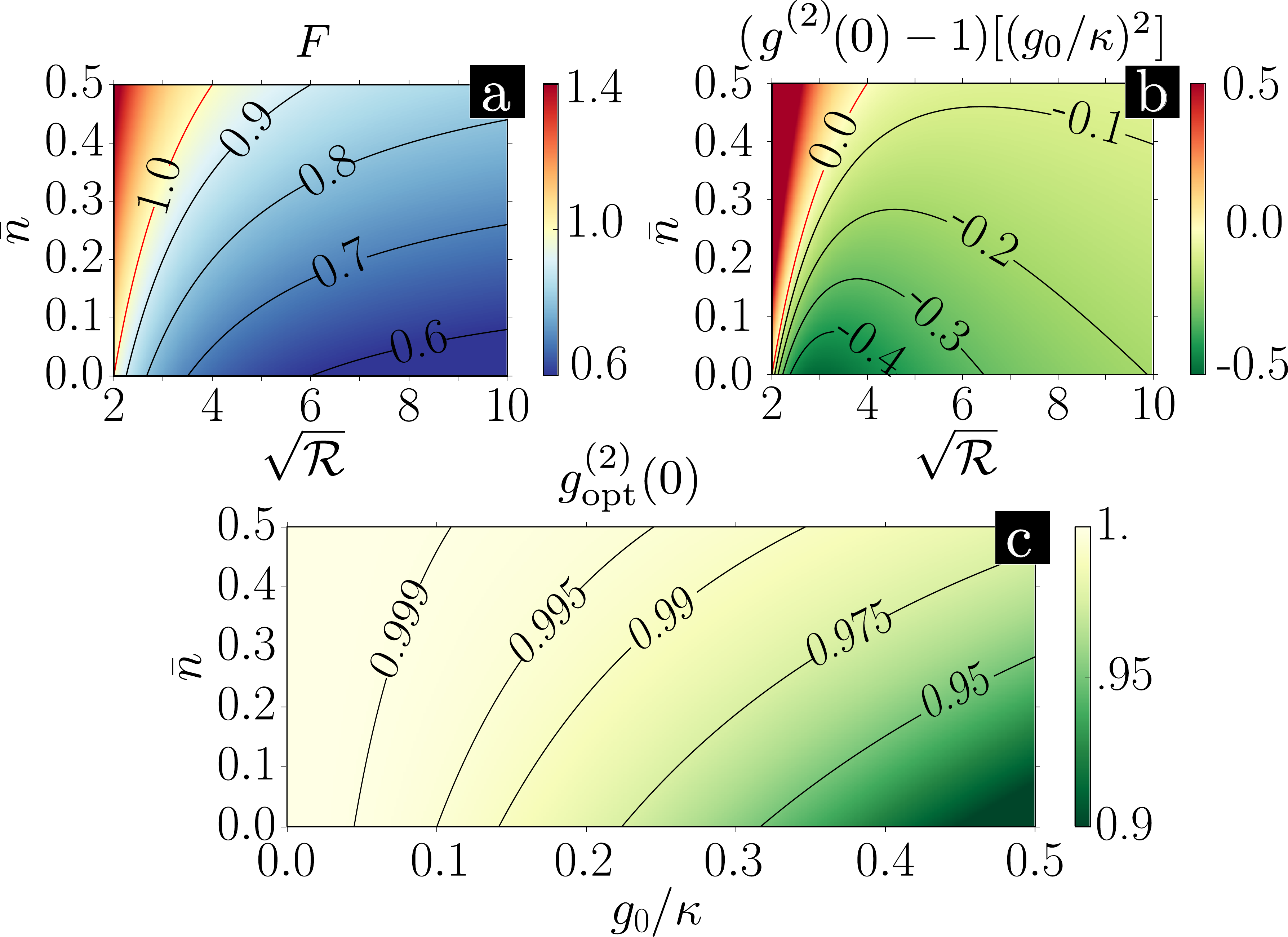}
	\caption{(a) Fano factor as a function of effective mechanical bath occupation number $\bar n$ and total number of photons in the cavity $n_{\mathrm{ph}}[{\kappa \gamma}/{4g_0^2}]= \sqrt{\mathcal R}$  according to Eqs.~ \eqref{Fano} and \eqref{nPH}. (b) Plot of $(g^{(2)}(0)-1)[(g_0/\kappa)^2]$ as a function of the same parameters in units of the squared single-photon strong-coupling parameter $(g_0/\kappa)^2$ according to Eqs.~ \eqref{g2} and \eqref{nPH}. The condition for both sub-Poissonian statistics ($F<1$) and antibunching ($g^{(2)}(0)<1$) is visualized by the red contour line  $\bar n = 1-2/\sqrt{\mathcal R}$ in both plots. (c) Plot of $g^{(2)}_\mathrm{opt}(0)$ for optimal choice of $\mathcal{R}$ as a function of $g_0/\kappa$ and $\bar n$ according to Eq.~ \eqref{optg2}.
	\label{FanoRequirement}
}
\end{figure}

\begin{comment}

\begin{figure}[t]
  \includegraphics[width=0.2\textwidth]{FanoRequirement.pdf}
	\caption{ Fano factor as a function of bare (!) laser power $\mathcal P_{B}$ and bare (!) bath occupation $\bar n_B$ with additional cooling laser operating at optimal value of $\gamma_{L}$. In this plot we use $n_{0}=0$ and allow also negative values of $\gamma_{L}$, which physically corresponds to operating the readout laser on the blue sideband. $\gamma_{L}<0$ can be useful to enhance $\mathcal C_{0}$ at the cost of $\bar n$. LABEL AUF BARE UMSTELLEN!
}
	\label{FanoRequirement2}
\end{figure}
\end{comment}

\paragraph{Additional Laser Cooling} Consider a setup where the mechanical oscillator is coupled to a third optical cavity of line width $\kappa_d$ which is driven below resonance such as to induce an additional damping $\gamma_{L}$ of the oscillator. Eliminating this cooling cavity gives rise to a `dressed' mechanical oscillator  whose equation of motion is still given by \eqref{Langevin} with an effective mechanical damping and occupation number
\begin{align}
\gamma &= \gamma_0+ \gamma_{L}, &
\bar n &= \frac{\gamma_0 \bar n_0 + \gamma_{L} \bar{n}_{L}}{\gamma_0+ \gamma_{L}}. \label{coolingTrafo}
\end{align}
Here $\gamma_0$ is the line width and $\bar{n}_0$ the occupation number of the bare mechanical resonance (without laser cooling), and $\bar{n}_L=(\kappa_d/4\omega_m)^2$ is the quantum limit of optomechanical laser cooling  \cite{WilsonRae2007,Marquardt2007}.

In order to have $F<1$ we assume laser cooling to an effective phonon occupation $\bar{n}<1$. This comes at the cost of a decreased gain parameter in Eq.~\eqref{gainR}, $\mathcal R={16 g_0^2 E^2}/{\kappa^3 (\gamma_0+\gamma_L)}$, which can be compensated for by a somewhat more intense driving field. It is rather remarkable that laser cooling can help to observe a quantum feature such as sub-Poissonian phonon statistics: While laser cooling can provide a small effective occupation number $\bar{n}\ll\bar{n}_0$ it does so by increasing the effective mechanical line width $\gamma\gg\gamma_0$ by the same factor. As a result, the decoherence rate
relevant for quantum effects, $\gamma_0\bar{n}_0=\gamma\bar{n}$, stays constant, such that laser cooling in most cases does not help in order to achieve quantum effects with mechanical oscillators.

\paragraph{Experimental feasibility with current technology} The requirements on the system parameters to have $g^{(2)}(0)<1$ (and therefore $F<1$) is found by inserting the mean amplitude \eqref{delta} and the Fano factor \eqref{Fano} in the definition \eqref{g2relF} of $g^{(2)}(0)$,
\begin{align}
&g^{(2)}(0)-1=  4\left(\frac{g_0}{\kappa}\right)^2 \frac {F-1} {\sqrt {\mathcal R}-1}. \label{g2}
\end{align}
For the discussion of experimental feasibility it is more instructive to express the gain parameter $\mathcal R$ in terms of  the steady state total number of photons in the cavity
 \begin{align}
n_{\mathrm{ph}}=|\alpha_0|^2+|\beta_0|^2= \frac {\kappa \gamma}{4g_0^2} \, \sqrt R, \label{nPH}
\end{align}
 where we used Eqs.~\eqref{delta} and \eqref{alphabeta}. The circulating number of photons is important as it determines the heating of the mechanical structure, which was the limiting decoherence mechanism in recent experiments with optomechanical crystals \cite{Chan2011a}.
In Fig.~\ref{FanoRequirement} (a) and (b) we show the Fano factor $F$ and $g^{(2)}(0)-1$ (in units of $g_0^2/\kappa^2$) as a function of
the number of photons in the cavity $n_{\mathrm{ph}}$ and the effective mechanical bath occupation number $\bar n$.
In view of the dependence of the Fano factor and the second order coherence function  on $\mathcal{R}$, cf. Eqs.~\eqref{Fano} and \eqref{g2} respectively, it is clear that there is an optimal number of circulating photons minimizing  $g^{(2)}(0)$ for given
 $\bar n$
 and single photon strong coupling parameter $g_0/\kappa$. The minimum is reached
 at $n_{\mathrm{ph}}\left[{\kappa \gamma}/{4g_0^2}\right]=({3+\bar n})/({1-\bar n})$ and is given by
\begin{align}
&g^{(2)}_\mathrm{opt}(0)=1-\tfrac 12 \left(\frac{g_0}{\kappa}\right)^2 \frac{(1-\bar n )^2}{(1+\bar n)},
\label{optg2}
\end{align}
which is illustrated in Fig. \ref{FanoRequirement}c.
Thus, a large single-photon coupling
 helps, but is not strictly required, to create a robust signal to verify antibunching.
We conclude that a sub-Poissonian phonon laser state can be prepared and verified outside the single-photon strong-coupling regime and for small but finite effective (cf. Eq.~ \eqref{coolingTrafo}) bath occupation $\bar n$ by detecting photon antibunching in the reflected light. We emphasize that phonon antibunching can be observed already in a regime of few circulating photons $n_{\mathrm{ph}} \ll 1$.

\paragraph{Readout}

The readout of the -- possibly antibunched -- phonon statistics can be
implemented in analogy to \cite{Cohen2014} using the cooling laser mode $d$. In the sideband resolved ($\kappa_d \ll \omega_m$)
  and linear ($g_d \zeta \ll \omega_m$)
 regime
the dynamics of laser cooling can be understood as a continuous coherent state swap interaction $c d^\dagger + c^\dagger d$ \cite{Verhagen2012,Palomaki2013}.
 The phonon statistics of $d$ can then be measured by counting the photons in the output of the cooling cavity $d$ at the sideband frequency $+\omega_m$ \cite{Cohen2014}. Hence with this readout scheme phonon antibunching is detected via photon antibunching.

\paragraph{Experimental case study}

Currently the highest reported value for the coupling in optomechanical crystals is $g_0/2 \pi=1.1$ MHz\cite{Chan2012}. The lowest cavity decay rate in a photonic crystal is, to our knowledge,
$\kappa =20$ MHz \cite{Sekoguchi2014}. While the best ratio achieved in a single device is $g_0/\kappa=0.007$ \cite{Chan2011a}, combining the best values in one device would already reach $g_0/\kappa \approx 0.055$. The lowest reported effective bath occupation reached with optomechanical cooling is $\bar n=0.85$ \cite{Chan2011a}, using a dilution refrigerator mechanical oscillators have even been cooled down below $\bar n<0.07$. Assuming a slightly more optimistic $g_0/\kappa=0.1$, an effective environmental temperature of 200 mK and a mechanical frequency of 5GHz  the deviation of $g^{(2)}_\mathrm{opt}(0)$ from 1 according to Eq.~ \eqref{optg2} will be 2.5 per mille. Further improvements on $g_0$ and $\kappa$ are expected using new designs and fabrication methods, so that reaching a signal of $g^{(2)}_\mathrm{opt}(0)-1$ on the order of a few per cent is a realistic prospect for the near future, cf. Fig.~ \ref{FanoRequirement}.
\paragraph{Outlook: Towards the single-photon strong-coupling regime}

Our linearized model is strictly valid only for $g_0/\kappa \ll 1$. We can however expect qualitative agreement to some extend even for larger $g_0/\kappa$. The deviation of $F$ from equation \eqref{Fano} in this regime are plotted in Fig. \ref{negWig}a. Strongly sub-Poissonian states with small limit cycle amplitude $\langle \hat n \rangle$ feature a negative Wigner function \cite{Lorch2014}. As discussed above $\langle \hat n \rangle \sim (\kappa/g_0)^2$ . It is therefore reasonable to expect negative mechanical Wigner density with $g_0/\kappa$ approaching the single-photon strong-coupling regime. Indeed we numerically find that negative Wigner density is possible for larger  $g_0/\kappa$ as depicted in Fig. \ref{negWig}b. All numerical calculations were done with QuTiP \cite{Johansson2011a,Johansson2013}, the details of the methods are discussed in the Appendix.

\paragraph{Conclusion}
Using an optomechanical setup with two optical modes brings experimental demonstration of both sub-Poissonian phonon statistics  and optomechanically induced phonon and photon antibunching in reach of today's technology. For parameters approaching the single-photon strong-coupling regime the limit cycle states can even feature a negative mechanical Wigner function.

\paragraph{Acknowledgements}
This work was funded by the Centre for QuantumEngineering and Space-Time Research (QUEST) at the Leibniz University of Hannover and by the European Community (FP7-Programme) through iQUOEMS (Grant Agreement No. 323924). We acknowledge the support of the cluster system team at the Leibniz University of Hannover in the production of this work.

\bibliographystyle{apsrev4-1}

\bibliography{/home/niloer/Documents/mendeleybibtex/TwoModeLimitCycles}

\begin{thebibliography}{49}%
\makeatletter
\providecommand \@ifxundefined [1]{%
 \@ifx{#1\undefined}
}%
\providecommand \@ifnum [1]{%
 \ifnum #1\expandafter \@firstoftwo
 \else \expandafter \@secondoftwo
 \fi
}%
\providecommand \@ifx [1]{%
 \ifx #1\expandafter \@firstoftwo
 \else \expandafter \@secondoftwo
 \fi
}%
\providecommand \natexlab [1]{#1}%
\providecommand \enquote  [1]{``#1''}%
\providecommand \bibnamefont  [1]{#1}%
\providecommand \bibfnamefont [1]{#1}%
\providecommand \citenamefont [1]{#1}%
\providecommand \href@noop [0]{\@secondoftwo}%
\providecommand \href [0]{\begingroup \@sanitize@url \@href}%
\providecommand \@href[1]{\@@startlink{#1}\@@href}%
\providecommand \@@href[1]{\endgroup#1\@@endlink}%
\providecommand \@sanitize@url [0]{\catcode `\\12\catcode `\$12\catcode
  `\&12\catcode `\#12\catcode `\^12\catcode `\_12\catcode `\%12\relax}%
\providecommand \@@startlink[1]{}%
\providecommand \@@endlink[0]{}%
\providecommand \url  [0]{\begingroup\@sanitize@url \@url }%
\providecommand \@url [1]{\endgroup\@href {#1}{\urlprefix }}%
\providecommand \urlprefix  [0]{URL }%
\providecommand \Eprint [0]{\href }%
\providecommand \doibase [0]{http://dx.doi.org/}%
\providecommand \selectlanguage [0]{\@gobble}%
\providecommand \bibinfo  [0]{\@secondoftwo}%
\providecommand \bibfield  [0]{\@secondoftwo}%
\providecommand \translation [1]{[#1]}%
\providecommand \BibitemOpen [0]{}%
\providecommand \bibitemStop [0]{}%
\providecommand \bibitemNoStop [0]{.\EOS\space}%
\providecommand \EOS [0]{\spacefactor3000\relax}%
\providecommand \BibitemShut  [1]{\csname bibitem#1\endcsname}%
\let\auto@bib@innerbib\@empty
%</preamble>
\bibitem [{\citenamefont {Aspelmeyer}\ \emph
  {et~al.}(2014{\natexlab{a}})\citenamefont {Aspelmeyer}, \citenamefont
  {Kippenberg},\ and\ \citenamefont {Marquardt}}]{Aspelmeyer2014}%
  \BibitemOpen
  \bibfield  {author} {\bibinfo {author} {\bibfnamefont {M.}~\bibnamefont
  {Aspelmeyer}}, \bibinfo {author} {\bibfnamefont {T.~J.}\ \bibnamefont
  {Kippenberg}}, \ and\ \bibinfo {author} {\bibfnamefont {F.}~\bibnamefont
  {Marquardt}},\ }\href {\doibase 10.1007/978-3-642-55312-7} {\emph {\bibinfo
  {title} {{Cavity Optomechanics}}}},\ edited by\ \bibinfo {editor}
  {\bibfnamefont {M.}~\bibnamefont {Aspelmeyer}}, \bibinfo {editor}
  {\bibfnamefont {T.~J.}\ \bibnamefont {Kippenberg}}, \ and\ \bibinfo {editor}
  {\bibfnamefont {F.}~\bibnamefont {Marquardt}}\ (\bibinfo  {publisher}
  {Springer Berlin Heidelberg},\ \bibinfo {address} {Berlin, Heidelberg},\
  \bibinfo {year} {2014})\BibitemShut {NoStop}%
\bibitem [{\citenamefont {Aspelmeyer}\ \emph
  {et~al.}(2014{\natexlab{b}})\citenamefont {Aspelmeyer}, \citenamefont
  {Kippenberg},\ and\ \citenamefont {Marquardt}}]{Aspelmeyer2014a}%
  \BibitemOpen
  \bibfield  {author} {\bibinfo {author} {\bibfnamefont {M.}~\bibnamefont
  {Aspelmeyer}}, \bibinfo {author} {\bibfnamefont {T.~J.}\ \bibnamefont
  {Kippenberg}}, \ and\ \bibinfo {author} {\bibfnamefont {F.}~\bibnamefont
  {Marquardt}},\ }\href {\doibase 10.1103/RevModPhys.86.1391} {\bibfield
  {journal} {\bibinfo  {journal} {Reviews of Modern Physics}\ }\textbf
  {\bibinfo {volume} {86}} (\bibinfo {year} {2014}{\natexlab{b}}),\
  10.1103/RevModPhys.86.1391},\ \Eprint {http://arxiv.org/abs/0712.1618}
  {arXiv:0712.1618} \BibitemShut {NoStop}%
\bibitem [{\citenamefont {Teufel}\ \emph {et~al.}(2011)\citenamefont {Teufel},
  \citenamefont {Donner}, \citenamefont {Li}, \citenamefont {Harlow},
  \citenamefont {Allman}, \citenamefont {Cicak}, \citenamefont {Sirois},
  \citenamefont {Whittaker}, \citenamefont {Lehnert},\ and\ \citenamefont
  {Simmonds}}]{Teufel2011}%
  \BibitemOpen
  \bibfield  {author} {\bibinfo {author} {\bibfnamefont {J.~D.}\ \bibnamefont
  {Teufel}}, \bibinfo {author} {\bibfnamefont {T.}~\bibnamefont {Donner}},
  \bibinfo {author} {\bibfnamefont {D.}~\bibnamefont {Li}}, \bibinfo {author}
  {\bibfnamefont {J.~W.}\ \bibnamefont {Harlow}}, \bibinfo {author}
  {\bibfnamefont {M.~S.}\ \bibnamefont {Allman}}, \bibinfo {author}
  {\bibfnamefont {K.}~\bibnamefont {Cicak}}, \bibinfo {author} {\bibfnamefont
  {A.~J.}\ \bibnamefont {Sirois}}, \bibinfo {author} {\bibfnamefont {J.~D.}\
  \bibnamefont {Whittaker}}, \bibinfo {author} {\bibfnamefont {K.~W.}\
  \bibnamefont {Lehnert}}, \ and\ \bibinfo {author} {\bibfnamefont {R.~W.}\
  \bibnamefont {Simmonds}},\ }\href {http://dx.doi.org/10.1038/nature10261}
  {\bibfield  {journal} {\bibinfo  {journal} {Nature}\ }\textbf {\bibinfo
  {volume} {475}},\ \bibinfo {pages} {359} (\bibinfo {year}
  {2011})}\BibitemShut {NoStop}%
\bibitem [{\citenamefont {Chan}\ \emph {et~al.}(2011)\citenamefont {Chan},
  \citenamefont {Alegre}, \citenamefont {Safavi-Naeini}, \citenamefont {Hill},
  \citenamefont {Krause}, \citenamefont {Gr\"{o}blacher}, \citenamefont
  {Aspelmeyer},\ and\ \citenamefont {Painter}}]{Chan2011a}%
  \BibitemOpen
  \bibfield  {author} {\bibinfo {author} {\bibfnamefont {J.}~\bibnamefont
  {Chan}}, \bibinfo {author} {\bibfnamefont {T.~P.~M.}\ \bibnamefont {Alegre}},
  \bibinfo {author} {\bibfnamefont {A.~H.}\ \bibnamefont {Safavi-Naeini}},
  \bibinfo {author} {\bibfnamefont {J.~T.}\ \bibnamefont {Hill}}, \bibinfo
  {author} {\bibfnamefont {A.}~\bibnamefont {Krause}}, \bibinfo {author}
  {\bibfnamefont {S.}~\bibnamefont {Gr\"{o}blacher}}, \bibinfo {author}
  {\bibfnamefont {M.}~\bibnamefont {Aspelmeyer}}, \ and\ \bibinfo {author}
  {\bibfnamefont {O.}~\bibnamefont {Painter}},\ }\href
  {http://dx.doi.org/10.1038/nature10461} {\bibfield  {journal} {\bibinfo
  {journal} {Nature}\ }\textbf {\bibinfo {volume} {478}},\ \bibinfo {pages}
  {89} (\bibinfo {year} {2011})}\BibitemShut {NoStop}%
\bibitem [{\citenamefont {Verhagen}\ \emph {et~al.}(2012)\citenamefont
  {Verhagen}, \citenamefont {Del\'{e}glise}, \citenamefont {Weis},
  \citenamefont {Schliesser},\ and\ \citenamefont {Kippenberg}}]{Verhagen2012}%
  \BibitemOpen
  \bibfield  {author} {\bibinfo {author} {\bibfnamefont {E.}~\bibnamefont
  {Verhagen}}, \bibinfo {author} {\bibfnamefont {S.}~\bibnamefont
  {Del\'{e}glise}}, \bibinfo {author} {\bibfnamefont {S.}~\bibnamefont {Weis}},
  \bibinfo {author} {\bibfnamefont {a.}~\bibnamefont {Schliesser}}, \ and\
  \bibinfo {author} {\bibfnamefont {T.~J.}\ \bibnamefont {Kippenberg}},\ }\href
  {\doibase 10.1038/nature10787} {\bibfield  {journal} {\bibinfo  {journal}
  {Nature}\ }\textbf {\bibinfo {volume} {482}},\ \bibinfo {pages} {63}
  (\bibinfo {year} {2012})}\BibitemShut {NoStop}%
\bibitem [{\citenamefont {Palomaki}\ \emph
  {et~al.}(2013{\natexlab{a}})\citenamefont {Palomaki}, \citenamefont {Harlow},
  \citenamefont {Teufel}, \citenamefont {Simmonds},\ and\ \citenamefont
  {Lehnert}}]{Palomaki2013}%
  \BibitemOpen
  \bibfield  {author} {\bibinfo {author} {\bibfnamefont {T.~A.}\ \bibnamefont
  {Palomaki}}, \bibinfo {author} {\bibfnamefont {J.~W.}\ \bibnamefont
  {Harlow}}, \bibinfo {author} {\bibfnamefont {J.~D.}\ \bibnamefont {Teufel}},
  \bibinfo {author} {\bibfnamefont {R.~W.}\ \bibnamefont {Simmonds}}, \ and\
  \bibinfo {author} {\bibfnamefont {K.~W.}\ \bibnamefont {Lehnert}},\ }\href
  {http://dx.doi.org/10.1038/nature11915} {\bibfield  {journal} {\bibinfo
  {journal} {Nature}\ }\textbf {\bibinfo {volume} {495}},\ \bibinfo {pages}
  {210} (\bibinfo {year} {2013}{\natexlab{a}})}\BibitemShut {NoStop}%
\bibitem [{\citenamefont {Murch}\ \emph {et~al.}(2008)\citenamefont {Murch},
  \citenamefont {Moore}, \citenamefont {Gupta},\ and\ \citenamefont
  {Stamper-Kurn}}]{Murch2008a}%
  \BibitemOpen
  \bibfield  {author} {\bibinfo {author} {\bibfnamefont {K.~W.}\ \bibnamefont
  {Murch}}, \bibinfo {author} {\bibfnamefont {K.~L.}\ \bibnamefont {Moore}},
  \bibinfo {author} {\bibfnamefont {S.}~\bibnamefont {Gupta}}, \ and\ \bibinfo
  {author} {\bibfnamefont {D.~M.}\ \bibnamefont {Stamper-Kurn}},\ }\href
  {http://dx.doi.org/10.1038/nphys965} {\bibfield  {journal} {\bibinfo
  {journal} {Nature Physics}\ }\textbf {\bibinfo {volume} {4}},\ \bibinfo
  {pages} {561} (\bibinfo {year} {2008})}\BibitemShut {NoStop}%
\bibitem [{\citenamefont {Purdy}\ \emph {et~al.}(2013)\citenamefont {Purdy},
  \citenamefont {Peterson},\ and\ \citenamefont {Regal}}]{Purdy2013a}%
  \BibitemOpen
  \bibfield  {author} {\bibinfo {author} {\bibfnamefont {T.~P.}\ \bibnamefont
  {Purdy}}, \bibinfo {author} {\bibfnamefont {R.~W.}\ \bibnamefont {Peterson}},
  \ and\ \bibinfo {author} {\bibfnamefont {C.~A.}\ \bibnamefont {Regal}},\
  }\href {\doibase 10.1126/science.1231282} {\bibfield  {journal} {\bibinfo
  {journal} {Science}\ }\textbf {\bibinfo {volume} {339}},\ \bibinfo {pages}
  {801} (\bibinfo {year} {2013})}\BibitemShut {NoStop}%
\bibitem [{\citenamefont {Palomaki}\ \emph
  {et~al.}(2013{\natexlab{b}})\citenamefont {Palomaki}, \citenamefont {Teufel},
  \citenamefont {Simmonds},\ and\ \citenamefont {Lehnert}}]{Palomaki2013a}%
  \BibitemOpen
  \bibfield  {author} {\bibinfo {author} {\bibfnamefont {T.~A.}\ \bibnamefont
  {Palomaki}}, \bibinfo {author} {\bibfnamefont {J.~D.}\ \bibnamefont
  {Teufel}}, \bibinfo {author} {\bibfnamefont {R.~W.}\ \bibnamefont
  {Simmonds}}, \ and\ \bibinfo {author} {\bibfnamefont {K.~W.}\ \bibnamefont
  {Lehnert}},\ }\href
  {http://www.sciencemag.org/content/early/2013/10/02/science.1244563}
  {\bibfield  {journal} {\bibinfo  {journal} {Science}\ }\textbf {\bibinfo
  {volume} {342}},\ \bibinfo {pages} {710} (\bibinfo {year}
  {2013}{\natexlab{b}})}\BibitemShut {NoStop}%
\bibitem [{\citenamefont {Carmon}\ \emph {et~al.}(2005)\citenamefont {Carmon},
  \citenamefont {Rokhsari}, \citenamefont {Yang}, \citenamefont {Kippenberg},\
  and\ \citenamefont {Vahala}}]{Carmon2005}%
  \BibitemOpen
  \bibfield  {author} {\bibinfo {author} {\bibfnamefont {T.}~\bibnamefont
  {Carmon}}, \bibinfo {author} {\bibfnamefont {H.}~\bibnamefont {Rokhsari}},
  \bibinfo {author} {\bibfnamefont {L.}~\bibnamefont {Yang}}, \bibinfo {author}
  {\bibfnamefont {T.}~\bibnamefont {Kippenberg}}, \ and\ \bibinfo {author}
  {\bibfnamefont {K.}~\bibnamefont {Vahala}},\ }\href
  {http://link.aps.org/doi/10.1103/PhysRevLett.94.223902} {\bibfield  {journal}
  {\bibinfo  {journal} {Physical Review Letters}\ }\textbf {\bibinfo {volume}
  {94}},\ \bibinfo {pages} {223902} (\bibinfo {year} {2005})}\BibitemShut
  {NoStop}%
\bibitem [{\citenamefont {Tomes}\ and\ \citenamefont
  {Carmon}(2009)}]{Tomes2009}%
  \BibitemOpen
  \bibfield  {author} {\bibinfo {author} {\bibfnamefont {M.}~\bibnamefont
  {Tomes}}\ and\ \bibinfo {author} {\bibfnamefont {T.}~\bibnamefont {Carmon}},\
  }\href {\doibase 10.1103/PhysRevLett.102.113601} {\bibfield  {journal}
  {\bibinfo  {journal} {Physical Review Letters}\ }\textbf {\bibinfo {volume}
  {102}},\ \bibinfo {pages} {20} (\bibinfo {year} {2009})}\BibitemShut
  {NoStop}%
\bibitem [{\citenamefont {Grudinin}\ \emph {et~al.}(2009)\citenamefont
  {Grudinin}, \citenamefont {Matsko},\ and\ \citenamefont
  {Maleki}}]{Grudinin2009}%
  \BibitemOpen
  \bibfield  {author} {\bibinfo {author} {\bibfnamefont {I.~S.}\ \bibnamefont
  {Grudinin}}, \bibinfo {author} {\bibfnamefont {A.~B.}\ \bibnamefont
  {Matsko}}, \ and\ \bibinfo {author} {\bibfnamefont {L.}~\bibnamefont
  {Maleki}},\ }\href {\doibase 10.1103/PhysRevLett.102.043902} {\bibfield
  {journal} {\bibinfo  {journal} {Physical Review Letters}\ }\textbf {\bibinfo
  {volume} {102}},\ \bibinfo {pages} {1} (\bibinfo {year} {2009})}\BibitemShut
  {NoStop}%
\bibitem [{\citenamefont {Grudinin}\ \emph {et~al.}(2010)\citenamefont
  {Grudinin}, \citenamefont {Lee}, \citenamefont {Painter},\ and\ \citenamefont
  {Vahala}}]{Grudinin2010}%
  \BibitemOpen
  \bibfield  {author} {\bibinfo {author} {\bibfnamefont {I.~S.}\ \bibnamefont
  {Grudinin}}, \bibinfo {author} {\bibfnamefont {H.}~\bibnamefont {Lee}},
  \bibinfo {author} {\bibfnamefont {O.}~\bibnamefont {Painter}}, \ and\
  \bibinfo {author} {\bibfnamefont {K.~J.}\ \bibnamefont {Vahala}},\ }\href
  {http://link.aps.org/doi/10.1103/PhysRevLett.104.083901} {\bibfield
  {journal} {\bibinfo  {journal} {Physical Review Letters}\ }\textbf {\bibinfo
  {volume} {104}},\ \bibinfo {pages} {083901} (\bibinfo {year}
  {2010})}\BibitemShut {NoStop}%
\bibitem [{\citenamefont {Bahl}\ \emph {et~al.}(2011)\citenamefont {Bahl},
  \citenamefont {Zehnpfennig}, \citenamefont {Tomes},\ and\ \citenamefont
  {Carmon}}]{Bahl2011}%
  \BibitemOpen
  \bibfield  {author} {\bibinfo {author} {\bibfnamefont {G.}~\bibnamefont
  {Bahl}}, \bibinfo {author} {\bibfnamefont {J.}~\bibnamefont {Zehnpfennig}},
  \bibinfo {author} {\bibfnamefont {M.}~\bibnamefont {Tomes}}, \ and\ \bibinfo
  {author} {\bibfnamefont {T.}~\bibnamefont {Carmon}},\ }\href {\doibase
  10.1038/ncomms1412} {\bibfield  {journal} {\bibinfo  {journal} {Nature
  communications}\ }\textbf {\bibinfo {volume} {2}},\ \bibinfo {pages} {403}
  (\bibinfo {year} {2011})},\ \Eprint {http://arxiv.org/abs/1106.2582}
  {arXiv:1106.2582} \BibitemShut {NoStop}%
\bibitem [{\citenamefont {Anetsberger}\ \emph {et~al.}(2011)\citenamefont
  {Anetsberger}, \citenamefont {Weig}, \citenamefont {Kotthaus},\ and\
  \citenamefont {Kippenberg}}]{Anetsberger2011}%
  \BibitemOpen
  \bibfield  {author} {\bibinfo {author} {\bibfnamefont {G.}~\bibnamefont
  {Anetsberger}}, \bibinfo {author} {\bibfnamefont {E.~M.}\ \bibnamefont
  {Weig}}, \bibinfo {author} {\bibfnamefont {J.~P.}\ \bibnamefont {Kotthaus}},
  \ and\ \bibinfo {author} {\bibfnamefont {T.~J.}\ \bibnamefont {Kippenberg}},\
  }\href {\doibase 10.1016/j.crhy.2011.10.012} {\bibfield  {journal} {\bibinfo
  {journal} {Comptes Rendus Physique}\ }\textbf {\bibinfo {volume} {12}},\
  \bibinfo {pages} {800} (\bibinfo {year} {2011})}\BibitemShut {NoStop}%
\bibitem [{\citenamefont {Zaitsev}\ \emph {et~al.}(2011)\citenamefont
  {Zaitsev}, \citenamefont {Pandey}, \citenamefont {Shtempluck},\ and\
  \citenamefont {Buks}}]{Zaitsev2011}%
  \BibitemOpen
  \bibfield  {author} {\bibinfo {author} {\bibfnamefont {S.}~\bibnamefont
  {Zaitsev}}, \bibinfo {author} {\bibfnamefont {A.~K.}\ \bibnamefont {Pandey}},
  \bibinfo {author} {\bibfnamefont {O.}~\bibnamefont {Shtempluck}}, \ and\
  \bibinfo {author} {\bibfnamefont {E.}~\bibnamefont {Buks}},\ }\href
  {http://link.aps.org/doi/10.1103/PhysRevE.84.046605} {\bibfield  {journal}
  {\bibinfo  {journal} {Physical Review E}\ }\textbf {\bibinfo {volume} {84}},\
  \bibinfo {pages} {046605} (\bibinfo {year} {2011})}\BibitemShut {NoStop}%
\bibitem [{\citenamefont {Suchoi}\ \emph {et~al.}(2014)\citenamefont {Suchoi},
  \citenamefont {Shlomi}, \citenamefont {Ella},\ and\ \citenamefont
  {Buks}}]{Suchoi}%
  \BibitemOpen
  \bibfield  {author} {\bibinfo {author} {\bibfnamefont {O.}~\bibnamefont
  {Suchoi}}, \bibinfo {author} {\bibfnamefont {K.}~\bibnamefont {Shlomi}},
  \bibinfo {author} {\bibfnamefont {L.}~\bibnamefont {Ella}}, \ and\ \bibinfo
  {author} {\bibfnamefont {E.}~\bibnamefont {Buks}},\ }\href
  {http://arxiv.org/abs/1408.2331} {\  (\bibinfo {year} {2014})},\ \Eprint
  {http://arxiv.org/abs/1408.2331v1} {arXiv:1408.2331v1} \BibitemShut {NoStop}%
\bibitem [{\citenamefont {Cohen}\ \emph {et~al.}(2014)\citenamefont {Cohen},
  \citenamefont {Meenehan}, \citenamefont {MacCabe}, \citenamefont
  {Gr\"{o}blacher}, \citenamefont {Safavi-Naeini}, \citenamefont {Marsili},
  \citenamefont {Shaw},\ and\ \citenamefont {Painter}}]{Cohen2014}%
  \BibitemOpen
  \bibfield  {author} {\bibinfo {author} {\bibfnamefont {J.~D.}\ \bibnamefont
  {Cohen}}, \bibinfo {author} {\bibfnamefont {S.~M.}\ \bibnamefont {Meenehan}},
  \bibinfo {author} {\bibfnamefont {G.~S.}\ \bibnamefont {MacCabe}}, \bibinfo
  {author} {\bibfnamefont {S.}~\bibnamefont {Gr\"{o}blacher}}, \bibinfo
  {author} {\bibfnamefont {A.~H.}\ \bibnamefont {Safavi-Naeini}}, \bibinfo
  {author} {\bibfnamefont {F.}~\bibnamefont {Marsili}}, \bibinfo {author}
  {\bibfnamefont {M.~D.}\ \bibnamefont {Shaw}}, \ and\ \bibinfo {author}
  {\bibfnamefont {O.}~\bibnamefont {Painter}},\ }\href
  {http://arxiv.org/abs/1410.1047v1} {\  (\bibinfo {year} {2014})},\ \Eprint
  {http://arxiv.org/abs/1410.1047} {arXiv:1410.1047} \BibitemShut {NoStop}%
\bibitem [{\citenamefont {Rodrigues}\ and\ \citenamefont
  {Armour}(2010)}]{Rodrigues2010}%
  \BibitemOpen
  \bibfield  {author} {\bibinfo {author} {\bibfnamefont {D.~A.}\ \bibnamefont
  {Rodrigues}}\ and\ \bibinfo {author} {\bibfnamefont {A.~D.}\ \bibnamefont
  {Armour}},\ }\href@noop {} {\bibfield  {journal} {\bibinfo  {journal} {Phys.
  Rev. Lett.}\ }\textbf {\bibinfo {volume} {104}} (\bibinfo {year}
  {2010})}\BibitemShut {NoStop}%
\bibitem [{\citenamefont {Armour}\ and\ \citenamefont
  {Rodrigues}(2012)}]{Armour2012a}%
  \BibitemOpen
  \bibfield  {author} {\bibinfo {author} {\bibfnamefont {A.~D.}\ \bibnamefont
  {Armour}}\ and\ \bibinfo {author} {\bibfnamefont {D.~A.}\ \bibnamefont
  {Rodrigues}},\ }\href {http://dx.doi.org/10.1016/j.crhy.2012.03.006}
  {\bibfield  {journal} {\bibinfo  {journal} {Comptes Rendus Physique}\
  }\textbf {\bibinfo {volume} {13}},\ \bibinfo {pages} {440} (\bibinfo {year}
  {2012})}\BibitemShut {NoStop}%
\bibitem [{\citenamefont {Qian}\ \emph {et~al.}(2012)\citenamefont {Qian},
  \citenamefont {Clerk}, \citenamefont {Hammerer},\ and\ \citenamefont
  {Marquardt}}]{Qian2012}%
  \BibitemOpen
  \bibfield  {author} {\bibinfo {author} {\bibfnamefont {J.}~\bibnamefont
  {Qian}}, \bibinfo {author} {\bibfnamefont {A.~A.}\ \bibnamefont {Clerk}},
  \bibinfo {author} {\bibfnamefont {K.}~\bibnamefont {Hammerer}}, \ and\
  \bibinfo {author} {\bibfnamefont {F.}~\bibnamefont {Marquardt}},\ }\href@noop
  {} {\bibfield  {journal} {\bibinfo  {journal} {Physical Review Letters}\
  }\textbf {\bibinfo {volume} {109}},\ \bibinfo {pages} {253601} (\bibinfo
  {year} {2012})}\BibitemShut {NoStop}%
\bibitem [{\citenamefont {Nation}(2013)}]{Nation2013}%
  \BibitemOpen
  \bibfield  {author} {\bibinfo {author} {\bibfnamefont {P.~D.}\ \bibnamefont
  {Nation}},\ }\href {\doibase 10.1103/PhysRevA.88.053828} {\bibfield
  {journal} {\bibinfo  {journal} {Physical Review A}\ }\textbf {\bibinfo
  {volume} {88}},\ \bibinfo {pages} {053828} (\bibinfo {year}
  {2013})}\BibitemShut {NoStop}%
\bibitem [{\citenamefont {L\"{o}rch}\ \emph {et~al.}(2014)\citenamefont
  {L\"{o}rch}, \citenamefont {Qian}, \citenamefont {Clerk}, \citenamefont
  {Marquardt},\ and\ \citenamefont {Hammerer}}]{Lorch2014}%
  \BibitemOpen
  \bibfield  {author} {\bibinfo {author} {\bibfnamefont {N.}~\bibnamefont
  {L\"{o}rch}}, \bibinfo {author} {\bibfnamefont {J.}~\bibnamefont {Qian}},
  \bibinfo {author} {\bibfnamefont {A.}~\bibnamefont {Clerk}}, \bibinfo
  {author} {\bibfnamefont {F.}~\bibnamefont {Marquardt}}, \ and\ \bibinfo
  {author} {\bibfnamefont {K.}~\bibnamefont {Hammerer}},\ }\href {\doibase
  10.1103/PhysRevX.4.011015} {\bibfield  {journal} {\bibinfo  {journal}
  {Physical Review X}\ }\textbf {\bibinfo {volume} {4}},\ \bibinfo {pages}
  {011015} (\bibinfo {year} {2014})}\BibitemShut {NoStop}%
\bibitem [{\citenamefont {Nation}\ \emph {et~al.}(2015)\citenamefont {Nation},
  \citenamefont {Johansson}, \citenamefont {Blencowe},\ and\ \citenamefont
  {Rimberg}}]{Nation2015}%
  \BibitemOpen
  \bibfield  {author} {\bibinfo {author} {\bibfnamefont {P.~D.}\ \bibnamefont
  {Nation}}, \bibinfo {author} {\bibfnamefont {J.~R.}\ \bibnamefont
  {Johansson}}, \bibinfo {author} {\bibfnamefont {M.~P.}\ \bibnamefont
  {Blencowe}}, \ and\ \bibinfo {author} {\bibfnamefont {A.~J.}\ \bibnamefont
  {Rimberg}},\ }\href {\doibase 10.1103/PhysRevE.91.013307} {\bibfield
  {journal} {\bibinfo  {journal} {Phys. Rev. E}\ }\textbf {\bibinfo {volume}
  {91}},\ \bibinfo {pages} {13307} (\bibinfo {year} {2015})}\BibitemShut
  {NoStop}%
\bibitem [{\citenamefont {Safavi-Naeini}\ and\ \citenamefont
  {Painter}(2011)}]{Safavi-Naeini2011}%
  \BibitemOpen
  \bibfield  {author} {\bibinfo {author} {\bibfnamefont {A.~H.}\ \bibnamefont
  {Safavi-Naeini}}\ and\ \bibinfo {author} {\bibfnamefont {O.}~\bibnamefont
  {Painter}},\ }\href {\doibase 10.1088/1367-2630/13/1/013017} {\bibfield
  {journal} {\bibinfo  {journal} {New Journal of Physics}\ }\textbf {\bibinfo
  {volume} {13}},\ \bibinfo {pages} {013017} (\bibinfo {year}
  {2011})}\BibitemShut {NoStop}%
\bibitem [{\citenamefont {Ludwig}\ \emph {et~al.}(2012)\citenamefont {Ludwig},
  \citenamefont {Safavi-Naeini}, \citenamefont {Painter},\ and\ \citenamefont
  {Marquardt}}]{Ludwig2012}%
  \BibitemOpen
  \bibfield  {author} {\bibinfo {author} {\bibfnamefont {M.}~\bibnamefont
  {Ludwig}}, \bibinfo {author} {\bibfnamefont {A.~H.}\ \bibnamefont
  {Safavi-Naeini}}, \bibinfo {author} {\bibfnamefont {O.}~\bibnamefont
  {Painter}}, \ and\ \bibinfo {author} {\bibfnamefont {F.}~\bibnamefont
  {Marquardt}},\ }\href {\doibase 10.1103/PhysRevLett.109.063601} {\bibfield
  {journal} {\bibinfo  {journal} {Physical Review Letters}\ }\textbf {\bibinfo
  {volume} {109}},\ \bibinfo {pages} {063601} (\bibinfo {year}
  {2012})}\BibitemShut {NoStop}%
\bibitem [{\citenamefont {Xu}\ \emph {et~al.}(2015)\citenamefont {Xu},
  \citenamefont {Gullans},\ and\ \citenamefont {Taylor}}]{Xu2014}%
  \BibitemOpen
  \bibfield  {author} {\bibinfo {author} {\bibfnamefont {X.}~\bibnamefont
  {Xu}}, \bibinfo {author} {\bibfnamefont {M.}~\bibnamefont {Gullans}}, \ and\
  \bibinfo {author} {\bibfnamefont {J.~M.}\ \bibnamefont {Taylor}},\ }\href
  {\doibase 10.1103/PhysRevA.91.013818} {\bibfield  {journal} {\bibinfo
  {journal} {Phys. Rev. A}\ }\textbf {\bibinfo {volume} {91}},\ \bibinfo
  {pages} {13818} (\bibinfo {year} {2015})}\BibitemShut {NoStop}%
\bibitem [{\citenamefont {K\'{o}m\'{a}r}\ \emph {et~al.}(2013)\citenamefont
  {K\'{o}m\'{a}r}, \citenamefont {Bennett}, \citenamefont {Stannigel},
  \citenamefont {Habraken}, \citenamefont {Rabl}, \citenamefont {Zoller},\ and\
  \citenamefont {Lukin}}]{Komar2013}%
  \BibitemOpen
  \bibfield  {author} {\bibinfo {author} {\bibfnamefont {P.}~\bibnamefont
  {K\'{o}m\'{a}r}}, \bibinfo {author} {\bibfnamefont {S.~D.}\ \bibnamefont
  {Bennett}}, \bibinfo {author} {\bibfnamefont {K.}~\bibnamefont {Stannigel}},
  \bibinfo {author} {\bibfnamefont {S.~J.~M.}\ \bibnamefont {Habraken}},
  \bibinfo {author} {\bibfnamefont {P.}~\bibnamefont {Rabl}}, \bibinfo {author}
  {\bibfnamefont {P.}~\bibnamefont {Zoller}}, \ and\ \bibinfo {author}
  {\bibfnamefont {M.~D.}\ \bibnamefont {Lukin}},\ }\href@noop {} {\bibfield
  {journal} {\bibinfo  {journal} {Physical Review A - Atomic, Molecular, and
  Optical Physics}\ }\textbf {\bibinfo {volume} {87}} (\bibinfo {year}
  {2013})}\BibitemShut {NoStop}%
\bibitem [{\citenamefont {Rabl}(2011)}]{Rabl2011}%
  \BibitemOpen
  \bibfield  {author} {\bibinfo {author} {\bibfnamefont {P.}~\bibnamefont
  {Rabl}},\ }\href {\doibase 10.1103/PhysRevLett.107.063601} {\bibfield
  {journal} {\bibinfo  {journal} {Physical Review Letters}\ }\textbf {\bibinfo
  {volume} {107}},\ \bibinfo {pages} {063601} (\bibinfo {year}
  {2011})}\BibitemShut {NoStop}%
\bibitem [{\citenamefont {Nunnenkamp}\ \emph {et~al.}(2011)\citenamefont
  {Nunnenkamp}, \citenamefont {B\o~rkje},\ and\ \citenamefont
  {Girvin}}]{Nunnenkamp2011}%
  \BibitemOpen
  \bibfield  {author} {\bibinfo {author} {\bibfnamefont {A.}~\bibnamefont
  {Nunnenkamp}}, \bibinfo {author} {\bibfnamefont {K.}~\bibnamefont
  {B\o~rkje}}, \ and\ \bibinfo {author} {\bibfnamefont {S.~M.}\ \bibnamefont
  {Girvin}},\ }\href {\doibase 10.1103/PhysRevLett.107.063602} {\bibfield
  {journal} {\bibinfo  {journal} {Physical Review Letters}\ }\textbf {\bibinfo
  {volume} {107}},\ \bibinfo {pages} {63602} (\bibinfo {year}
  {2011})}\BibitemShut {NoStop}%
\bibitem [{\citenamefont {Braginsky}\ \emph {et~al.}(2001)\citenamefont
  {Braginsky}, \citenamefont {Strigin},\ and\ \citenamefont
  {Vyatchanin}}]{Braginsky2001}%
  \BibitemOpen
  \bibfield  {author} {\bibinfo {author} {\bibfnamefont {V.}~\bibnamefont
  {Braginsky}}, \bibinfo {author} {\bibfnamefont {S.}~\bibnamefont {Strigin}},
  \ and\ \bibinfo {author} {\bibfnamefont {S.}~\bibnamefont {Vyatchanin}},\
  }\href {http://dx.doi.org/10.1016/S0375-9601(01)00510-2} {\bibfield
  {journal} {\bibinfo  {journal} {Physics Letters A}\ }\textbf {\bibinfo
  {volume} {287}},\ \bibinfo {pages} {331} (\bibinfo {year}
  {2001})}\BibitemShut {NoStop}%
\bibitem [{\citenamefont {Chen}\ \emph {et~al.}(2014)\citenamefont {Chen},
  \citenamefont {C.Zhao}, \citenamefont {Danilishin}, \citenamefont {{L. Ju}},
  \citenamefont {Wang}, \citenamefont {Vyatchanin}, \citenamefont {Molinelli},
  \citenamefont {Kuhn}, \citenamefont {Gras}, \citenamefont {Briant},
  \citenamefont {Cohadon}, \citenamefont {Heidmann}, \citenamefont
  {Roch-Jeune}, \citenamefont {Flaminio}, \citenamefont {Michel},\ and\
  \citenamefont {Pinard}}]{Chen2014}%
  \BibitemOpen
  \bibfield  {author} {\bibinfo {author} {\bibfnamefont {X.}~\bibnamefont
  {Chen}}, \bibinfo {author} {\bibnamefont {C.Zhao}}, \bibinfo {author}
  {\bibfnamefont {S.}~\bibnamefont {Danilishin}}, \bibinfo {author}
  {\bibfnamefont {D.~B.}\ \bibnamefont {{L. Ju}}}, \bibinfo {author}
  {\bibfnamefont {H.}~\bibnamefont {Wang}}, \bibinfo {author} {\bibfnamefont
  {S.~P.}\ \bibnamefont {Vyatchanin}}, \bibinfo {author} {\bibfnamefont
  {C.}~\bibnamefont {Molinelli}}, \bibinfo {author} {\bibfnamefont
  {A.}~\bibnamefont {Kuhn}}, \bibinfo {author} {\bibfnamefont {S.}~\bibnamefont
  {Gras}}, \bibinfo {author} {\bibfnamefont {T.}~\bibnamefont {Briant}},
  \bibinfo {author} {\bibfnamefont {P.-F.}\ \bibnamefont {Cohadon}}, \bibinfo
  {author} {\bibfnamefont {A.}~\bibnamefont {Heidmann}}, \bibinfo {author}
  {\bibfnamefont {I.}~\bibnamefont {Roch-Jeune}}, \bibinfo {author}
  {\bibfnamefont {R.}~\bibnamefont {Flaminio}}, \bibinfo {author}
  {\bibfnamefont {C.}~\bibnamefont {Michel}}, \ and\ \bibinfo {author}
  {\bibfnamefont {L.}~\bibnamefont {Pinard}},\ }\href
  {http://arxiv.org/abs/1303.4561} {\  (\bibinfo {year} {2014})},\ \Eprint
  {http://arxiv.org/abs/1411.3016} {arXiv:1411.3016} \BibitemShut {NoStop}%
\bibitem [{\citenamefont {Wu}\ \emph {et~al.}(2013)\citenamefont {Wu},
  \citenamefont {Heinrich},\ and\ \citenamefont {Marquardt}}]{Wu2013}%
  \BibitemOpen
  \bibfield  {author} {\bibinfo {author} {\bibfnamefont {H.}~\bibnamefont
  {Wu}}, \bibinfo {author} {\bibfnamefont {G.}~\bibnamefont {Heinrich}}, \ and\
  \bibinfo {author} {\bibfnamefont {F.}~\bibnamefont {Marquardt}},\ }\href
  {\doibase 10.1088/1367-2630/15/12/123022} {\bibfield  {journal} {\bibinfo
  {journal} {New Journal of Physics}\ }\textbf {\bibinfo {volume} {15}},\
  \bibinfo {pages} {1} (\bibinfo {year} {2013})},\ \Eprint
  {http://arxiv.org/abs/1102.1647} {arXiv:1102.1647} \BibitemShut {NoStop}%
\bibitem [{\citenamefont {Danilishin}\ \emph {et~al.}(2014)\citenamefont
  {Danilishin}, \citenamefont {Vyatchanin}, \citenamefont {Blair},
  \citenamefont {Li},\ and\ \citenamefont {Zhao}}]{Danilishin2014}%
  \BibitemOpen
  \bibfield  {author} {\bibinfo {author} {\bibfnamefont {S.~L.}\ \bibnamefont
  {Danilishin}}, \bibinfo {author} {\bibfnamefont {S.~P.}\ \bibnamefont
  {Vyatchanin}}, \bibinfo {author} {\bibfnamefont {D.~G.}\ \bibnamefont
  {Blair}}, \bibinfo {author} {\bibfnamefont {J.}~\bibnamefont {Li}}, \ and\
  \bibinfo {author} {\bibfnamefont {C.}~\bibnamefont {Zhao}},\ }\href {\doibase
  10.1103/PhysRevD.90.122008} {\bibfield  {journal} {\bibinfo  {journal} {Phys.
  Rev. D}\ }\textbf {\bibinfo {volume} {90}},\ \bibinfo {pages} {122008}
  (\bibinfo {year} {2014})}\BibitemShut {NoStop}%
\bibitem [{\citenamefont {Ju}\ \emph {et~al.}(2014)\citenamefont {Ju},
  \citenamefont {Zhao}, \citenamefont {Ma}, \citenamefont {Blair},
  \citenamefont {Danilishin},\ and\ \citenamefont {Gras}}]{Ju2014}%
  \BibitemOpen
  \bibfield  {author} {\bibinfo {author} {\bibfnamefont {L.}~\bibnamefont
  {Ju}}, \bibinfo {author} {\bibfnamefont {C.}~\bibnamefont {Zhao}}, \bibinfo
  {author} {\bibfnamefont {Y.}~\bibnamefont {Ma}}, \bibinfo {author}
  {\bibfnamefont {D.}~\bibnamefont {Blair}}, \bibinfo {author} {\bibfnamefont
  {S.~L.}\ \bibnamefont {Danilishin}}, \ and\ \bibinfo {author} {\bibfnamefont
  {S.}~\bibnamefont {Gras}},\ }\href
  {http://stacks.iop.org/0264-9381/31/i=14/a=145002} {\bibfield  {journal}
  {\bibinfo  {journal} {Classical and Quantum Gravity}\ }\textbf {\bibinfo
  {volume} {31}},\ \bibinfo {pages} {145002} (\bibinfo {year}
  {2014})}\BibitemShut {NoStop}%
\bibitem [{\citenamefont {Wang}\ \emph {et~al.}(2014)\citenamefont {Wang},
  \citenamefont {Wang}, \citenamefont {Zhang}, \citenamefont {\"{O}zdemir},
  \citenamefont {Yang},\ and\ \citenamefont {Liu}}]{Wang2014a}%
  \BibitemOpen
  \bibfield  {author} {\bibinfo {author} {\bibfnamefont {H.}~\bibnamefont
  {Wang}}, \bibinfo {author} {\bibfnamefont {Z.}~\bibnamefont {Wang}}, \bibinfo
  {author} {\bibfnamefont {J.}~\bibnamefont {Zhang}}, \bibinfo {author}
  {\bibfnamefont {S.~K.}\ \bibnamefont {\"{O}zdemir}}, \bibinfo {author}
  {\bibfnamefont {L.}~\bibnamefont {Yang}}, \ and\ \bibinfo {author}
  {\bibfnamefont {Y.-X.}\ \bibnamefont {Liu}},\ }\href {\doibase
  10.1103/PhysRevA.90.053814} {\bibfield  {journal} {\bibinfo  {journal}
  {Physical Review A}\ }\textbf {\bibinfo {volume} {90}},\ \bibinfo {pages}
  {053814} (\bibinfo {year} {2014})}\BibitemShut {NoStop}%
\bibitem [{\citenamefont {Kimble}\ \emph {et~al.}(1977)\citenamefont {Kimble},
  \citenamefont {Dagenais},\ and\ \citenamefont {Mandel}}]{Kimble1977}%
  \BibitemOpen
  \bibfield  {author} {\bibinfo {author} {\bibfnamefont {H.~J.}\ \bibnamefont
  {Kimble}}, \bibinfo {author} {\bibfnamefont {M.}~\bibnamefont {Dagenais}}, \
  and\ \bibinfo {author} {\bibfnamefont {L.}~\bibnamefont {Mandel}},\
  }\href@noop {} {\bibfield  {journal} {\bibinfo  {journal} {Physical Review
  Letters}\ }\textbf {\bibinfo {volume} {39}},\ \bibinfo {pages} {691}
  (\bibinfo {year} {1977})}\BibitemShut {NoStop}%
\bibitem [{App()}]{Appendix}%
  \BibitemOpen
  \href@noop {} {\enquote {\bibinfo {title} {{see Appendix}},}\ }\BibitemShut
  {NoStop}%
\bibitem [{\citenamefont {Safavi-Naeini}\ \emph {et~al.}(2014)\citenamefont
  {Safavi-Naeini}, \citenamefont {Hill}, \citenamefont {Meenehan},
  \citenamefont {Chan}, \citenamefont {Gr\"{o}blacher},\ and\ \citenamefont
  {Painter}}]{Safavi-Naeini2014}%
  \BibitemOpen
  \bibfield  {author} {\bibinfo {author} {\bibfnamefont {A.~H.}\ \bibnamefont
  {Safavi-Naeini}}, \bibinfo {author} {\bibfnamefont {J.~T.}\ \bibnamefont
  {Hill}}, \bibinfo {author} {\bibfnamefont {S.}~\bibnamefont {Meenehan}},
  \bibinfo {author} {\bibfnamefont {J.}~\bibnamefont {Chan}}, \bibinfo {author}
  {\bibfnamefont {S.}~\bibnamefont {Gr\"{o}blacher}}, \ and\ \bibinfo {author}
  {\bibfnamefont {O.}~\bibnamefont {Painter}},\ }\href {\doibase
  10.1103/PhysRevLett.112.153603} {\bibfield  {journal} {\bibinfo  {journal}
  {Physical Review Letters}\ }\textbf {\bibinfo {volume} {112}},\ \bibinfo
  {pages} {153603} (\bibinfo {year} {2014})}\BibitemShut {NoStop}%
\bibitem [{\citenamefont {Meenehan}\ \emph {et~al.}(2014)\citenamefont
  {Meenehan}, \citenamefont {Cohen}, \citenamefont {Gr\"{o}blacher},
  \citenamefont {Hill}, \citenamefont {Safavi-Naeini}, \citenamefont
  {Aspelmeyer},\ and\ \citenamefont {Painter}}]{Meenehan2014}%
  \BibitemOpen
  \bibfield  {author} {\bibinfo {author} {\bibfnamefont {S.~M.}\ \bibnamefont
  {Meenehan}}, \bibinfo {author} {\bibfnamefont {J.~D.}\ \bibnamefont {Cohen}},
  \bibinfo {author} {\bibfnamefont {S.}~\bibnamefont {Gr\"{o}blacher}},
  \bibinfo {author} {\bibfnamefont {J.~T.}\ \bibnamefont {Hill}}, \bibinfo
  {author} {\bibfnamefont {A.~H.}\ \bibnamefont {Safavi-Naeini}}, \bibinfo
  {author} {\bibfnamefont {M.}~\bibnamefont {Aspelmeyer}}, \ and\ \bibinfo
  {author} {\bibfnamefont {O.}~\bibnamefont {Painter}},\ }\href {\doibase
  10.1103/PhysRevA.90.011803} {\bibfield  {journal} {\bibinfo  {journal} {Phys.
  Rev. A}\ }\textbf {\bibinfo {volume} {90}},\ \bibinfo {pages} {11803}
  (\bibinfo {year} {2014})}\BibitemShut {NoStop}%
\bibitem [{\citenamefont {Wilson-Rae}\ \emph {et~al.}(2007)\citenamefont
  {Wilson-Rae}, \citenamefont {Nooshi}, \citenamefont {Zwerger},\ and\
  \citenamefont {Kippenberg}}]{WilsonRae2007}%
  \BibitemOpen
  \bibfield  {author} {\bibinfo {author} {\bibfnamefont {I.}~\bibnamefont
  {Wilson-Rae}}, \bibinfo {author} {\bibfnamefont {N.}~\bibnamefont {Nooshi}},
  \bibinfo {author} {\bibfnamefont {W.}~\bibnamefont {Zwerger}}, \ and\
  \bibinfo {author} {\bibfnamefont {T.}~\bibnamefont {Kippenberg}},\ }\href
  {http://prl.aps.org/abstract/PRL/v99/i9/e093901} {\bibfield  {journal}
  {\bibinfo  {journal} {Physical Review Letters}\ }\textbf {\bibinfo {volume}
  {99}},\ \bibinfo {pages} {093901} (\bibinfo {year} {2007})}\BibitemShut
  {NoStop}%
\bibitem [{\citenamefont {Marquardt}\ \emph {et~al.}(2007)\citenamefont
  {Marquardt}, \citenamefont {Chen}, \citenamefont {Clerk},\ and\ \citenamefont
  {Girvin}}]{Marquardt2007}%
  \BibitemOpen
  \bibfield  {author} {\bibinfo {author} {\bibfnamefont {F.}~\bibnamefont
  {Marquardt}}, \bibinfo {author} {\bibfnamefont {J.}~\bibnamefont {Chen}},
  \bibinfo {author} {\bibfnamefont {A.}~\bibnamefont {Clerk}}, \ and\ \bibinfo
  {author} {\bibfnamefont {S.}~\bibnamefont {Girvin}},\ }\href
  {http://link.aps.org/doi/10.1103/PhysRevLett.99.093902} {\bibfield  {journal}
  {\bibinfo  {journal} {Physical Review Letters}\ }\textbf {\bibinfo {volume}
  {99}},\ \bibinfo {pages} {093902} (\bibinfo {year} {2007})}\BibitemShut
  {NoStop}%
\bibitem [{\citenamefont {Chan}\ \emph {et~al.}(2012)\citenamefont {Chan},
  \citenamefont {Safavi-Naeini}, \citenamefont {Hill}, \citenamefont
  {Meenehan},\ and\ \citenamefont {Painter}}]{Chan2012}%
  \BibitemOpen
  \bibfield  {author} {\bibinfo {author} {\bibfnamefont {J.}~\bibnamefont
  {Chan}}, \bibinfo {author} {\bibfnamefont {A.~H.}\ \bibnamefont
  {Safavi-Naeini}}, \bibinfo {author} {\bibfnamefont {J.~T.}\ \bibnamefont
  {Hill}}, \bibinfo {author} {\bibfnamefont {S.}~\bibnamefont {Meenehan}}, \
  and\ \bibinfo {author} {\bibfnamefont {O.}~\bibnamefont {Painter}},\ }\href
  {\doibase 10.1063/1.4747726} {\bibfield  {journal} {\bibinfo  {journal}
  {Applied Physics Letters}\ }\textbf {\bibinfo {volume} {101}},\ \bibinfo
  {pages} {081115} (\bibinfo {year} {2012})}\BibitemShut {NoStop}%
\bibitem [{\citenamefont {Sekoguchi}\ \emph {et~al.}(2014)\citenamefont
  {Sekoguchi}, \citenamefont {Takahashi}, \citenamefont {Asano},\ and\
  \citenamefont {Noda}}]{Sekoguchi2014}%
  \BibitemOpen
  \bibfield  {author} {\bibinfo {author} {\bibfnamefont {H.}~\bibnamefont
  {Sekoguchi}}, \bibinfo {author} {\bibfnamefont {Y.}~\bibnamefont
  {Takahashi}}, \bibinfo {author} {\bibfnamefont {T.}~\bibnamefont {Asano}}, \
  and\ \bibinfo {author} {\bibfnamefont {S.}~\bibnamefont {Noda}},\ }\href
  {\doibase 10.1364/OE.22.000916} {\bibfield  {journal} {\bibinfo  {journal}
  {Optics Express}\ }\textbf {\bibinfo {volume} {22}},\ \bibinfo {pages} {916}
  (\bibinfo {year} {2014})}\BibitemShut {NoStop}%
\bibitem [{\citenamefont {Johansson}\ \emph {et~al.}(2011)\citenamefont
  {Johansson}, \citenamefont {Nation},\ and\ \citenamefont
  {Nori}}]{Johansson2011a}%
  \BibitemOpen
  \bibfield  {author} {\bibinfo {author} {\bibfnamefont {J.~R.}\ \bibnamefont
  {Johansson}}, \bibinfo {author} {\bibfnamefont {P.~D.}\ \bibnamefont
  {Nation}}, \ and\ \bibinfo {author} {\bibfnamefont {F.}~\bibnamefont
  {Nori}},\ }\href {\doibase 10.1016/j.cpc.2012.02.021} {\bibfield  {journal}
  {\bibinfo  {journal} {Computer Physics Communications}\ }\textbf {\bibinfo
  {volume} {183}},\ \bibinfo {pages} {16} (\bibinfo {year} {2011})},\ \Eprint
  {http://arxiv.org/abs/1110.0573} {arXiv:1110.0573} \BibitemShut {NoStop}%
\bibitem [{\citenamefont {Johansson}\ \emph {et~al.}(2013)\citenamefont
  {Johansson}, \citenamefont {Nation},\ and\ \citenamefont
  {Nori}}]{Johansson2013}%
  \BibitemOpen
  \bibfield  {author} {\bibinfo {author} {\bibfnamefont {J.}~\bibnamefont
  {Johansson}}, \bibinfo {author} {\bibfnamefont {P.}~\bibnamefont {Nation}}, \
  and\ \bibinfo {author} {\bibfnamefont {F.}~\bibnamefont {Nori}},\ }\href
  {\doibase 10.1016/j.cpc.2012.11.019} {\bibfield  {journal} {\bibinfo
  {journal} {Computer Physics Communications}\ }\textbf {\bibinfo {volume}
  {184}},\ \bibinfo {pages} {1234} (\bibinfo {year} {2013})}\BibitemShut
  {NoStop}%
\bibitem [{\citenamefont {Dum}\ \emph {et~al.}(1992)\citenamefont {Dum},
  \citenamefont {Zoller},\ and\ \citenamefont {Ritsch}}]{Dum1992}%
  \BibitemOpen
  \bibfield  {author} {\bibinfo {author} {\bibfnamefont {R.}~\bibnamefont
  {Dum}}, \bibinfo {author} {\bibfnamefont {P.}~\bibnamefont {Zoller}}, \ and\
  \bibinfo {author} {\bibfnamefont {H.}~\bibnamefont {Ritsch}},\ }\href
  {\doibase 10.1103/PhysRevA.45.4879} {\bibfield  {journal} {\bibinfo
  {journal} {Physical Review A}\ }\textbf {\bibinfo {volume} {45}},\ \bibinfo
  {pages} {4879} (\bibinfo {year} {1992})}\BibitemShut {NoStop}%
\bibitem [{\citenamefont {Dalibard}\ \emph {et~al.}(1992)\citenamefont
  {Dalibard}, \citenamefont {Castin},\ and\ \citenamefont
  {M\o~lmer}}]{Dalibard1992}%
  \BibitemOpen
  \bibfield  {author} {\bibinfo {author} {\bibfnamefont {J.}~\bibnamefont
  {Dalibard}}, \bibinfo {author} {\bibfnamefont {Y.}~\bibnamefont {Castin}}, \
  and\ \bibinfo {author} {\bibfnamefont {K.}~\bibnamefont {M\o~lmer}},\ }\href
  {\doibase 10.1103/PhysRevLett.68.580} {\bibfield  {journal} {\bibinfo
  {journal} {Physical Review Letters}\ }\textbf {\bibinfo {volume} {68}},\
  \bibinfo {pages} {580} (\bibinfo {year} {1992})}\BibitemShut {NoStop}%
\bibitem [{\citenamefont {M\o~lmer}\ \emph {et~al.}(1993)\citenamefont
  {M\o~lmer}, \citenamefont {Castin},\ and\ \citenamefont
  {Dalibard}}]{Molmer1993}%
  \BibitemOpen
  \bibfield  {author} {\bibinfo {author} {\bibfnamefont {K.}~\bibnamefont
  {M\o~lmer}}, \bibinfo {author} {\bibfnamefont {Y.}~\bibnamefont {Castin}}, \
  and\ \bibinfo {author} {\bibfnamefont {J.}~\bibnamefont {Dalibard}},\ }\href
  {\doibase 10.1364/JOSAB.10.000524} {\bibfield  {journal} {\bibinfo  {journal}
  {Journal of the Optical Society of America B}\ }\textbf {\bibinfo {volume}
  {10}},\ \bibinfo {pages} {524} (\bibinfo {year} {1993})}\BibitemShut
  {NoStop}%
\end{thebibliography}%
%\bibliography{/Users/klhamm/Documents/Projekte/Bibliographies/TwoModeLimitCycles}
%\bibliography{TwoModeInstability6}

%\newpage

\appendix

\paragraph{numerics}
The steady state of the system was calculated using QuTiP \cite{Johansson2011a,Johansson2013}. For Fig.~ \ref{negWig}b, where the mechanical amplitudes  are small due to the large $g_0/\kappa$, the Hilbert space has moderate size and we used a direct steady state solver for density matrices. For Fig.~ \ref{negWig}a the Hilbert space is (in general) too large for this and we had to use Monte-Carlo trajectories \cite{Dum1992,Dalibard1992,Molmer1993} for the wave function and average over many runs to obtain a density matrix. Each trajectory $\ket {\psi_j(t)}$ had a coherent state with random, independent and identically distributed Gaussian amplitudes, $\xi_j \sim \mathcal N(\zeta,1)$  around the analytical steady state amplitude $\zeta$ from Eq.~ $\eqref{delta}$ as initial state for the oscillator and vacuum as initial state for both optical modes. The system was then evolved for a time $\tau=5/\gamma$ with Hamiltonian \eqref{totH} and the Lindblad operator $L=L_c+L_m$, where $L_c \rho=\kappa a \rho a^\dagger -\tfrac \kappa 2 a^\dagger a \rho - \tfrac \kappa 2 \rho a^\dagger a +\kappa b \rho b^\dagger -\tfrac \kappa 2 b^\dagger b \rho - \tfrac \kappa 2 \rho b^\dagger b$ and $L_m=\gamma c \rho c^\dagger -\tfrac \gamma 2 c^\dagger c \rho - \tfrac \gamma 2 \rho c^\dagger c$. The calculation was done in a displaced frame around the mean amplitude of the mechanical oscillator and cavity modes.
We then used $\sigma= \sum_j \ket { {\psi_j(\tau)}} \bra { {\psi_j(\tau)}}$ to calculate $\langle \hat n \rangle_\sigma$ and $\langle \hat n^2 \rangle_\sigma$, which in turn gives the Fano factor $F$ and $g^{(2)}(0)$. The
time $\tau$ was chosen such that both mean values had already relaxed to steady state, compare the damping in equation \eqref{quadratureEquation}, while the phase has still not diffused away too far from $\zeta$ so that a small Hilbert space around the mean mechanical amplitude was sufficient for the simulation.

\end{document}